\documentclass[acmsmall, nonacm]{acmart}

\usepackage{subfiles}
\usepackage{makecell}
\usepackage{enumitem}
\usepackage{multirow}
\usepackage{longtable}
\usepackage{graphicx}
\usepackage{subcaption}
\usepackage{listings}
\usepackage{pgfplots}
\usepackage{pgf-pie}
\usepackage{pgf-pie}
\usepackage{soul}
\pgfplotsset{compat=1.17}
\usepackage{amsmath}
\usepackage{array}
\usepackage{float}
\usepackage{pdflscape}
\usepackage{rotating}
\usepackage{xcolor}
\usepackage{csquotes}
\usepackage{booktabs}
\usepackage{tabularx}
\usepackage{longtable}

\newcolumntype{L}[1]{>{\raggedright\arraybackslash}p{#1}}
\definecolor{disagree}{HTML}{C0392B}
\newcommand{\disagree}[1]{\textcolor{disagree}{#1}}

\newif\ifredact
\redacttrue   

\newif\ifcomment
\commenttrue   
\ifcomment
  \newcommand{\missing}[1]{\textcolor{red}{~#1}}
  \newcommand{\kel}[1]{\sethlcolor{pink}\hl{~[Kellie: #1]}}
  \newcommand{\ken}[1]{\sethlcolor{yellow}\hl{~[Kenny: #1]}}
\else
  \newcommand{\missing}[1]{}
  \newcommand{\kel}[1]{}
  \newcommand{\ken}[1]{}
\fi

\newcommand{\dquote}[1]{\enquote{#1}}
\newcommand{\squote}[1]{\enquote*{#1}}

\AtBeginDocument{%
  }
\begin{document}
\renewcommand\footnotetextcopyrightpermission[1]{} 


\newcommand{\duskytide}{\textbf{\textit{PS1}}}
\newcommand{\zoomplane}{\textbf{\textit{PS2}}}
\newcommand{\windsong}{\textbf{\textit{PS3}}}
\newcommand{\amberlight}{\textbf{\textit{PS4}}}
\newcommand{\brightstar}{\textbf{\textit{PS5}}}
\newcommand{\butterflywave}{\textbf{\textit{PS6}}}
\newcommand{\bluemoon}{\textbf{\textit{PS7}}}
\newcommand{\greenship}{\textbf{\textit{PS8}}}
\newcommand{\ironpetal}{\textbf{\textit{PS9}}}
\newcommand{\earnprom}{\textbf{\textit{PS10}}}
\newcommand{\redballoon}{\textbf{\textit{PS11}}}
\newcommand{\greenleaf}{\textbf{\textit{PS12}}}

\newcommand{\stormvale}{\textbf{\textit{E1}}}
\newcommand{\velvetecho}{\textbf{\textit{E2}}}
\newcommand{\azureflame}{\textbf{\textit{E3}}}
\newcommand{\twilightrush}{\textbf{\textit{E4}}}
\newcommand{\mistybloom}{\textbf{\textit{E5}}}
\newcommand{\solarwisp}{\textbf{\textit{E6}}}

\newcommand{\empathai}{\textsc{EmpathAI}}
\newcommand{\client}{\textsc{SimClient}}
\newcommand{\suggestions}{\textsc{Suggestions}}
\newcommand{\smi}{\(S_{MI}\)}
\newcommand{\ser}{\(S_{ER}\)}
\newcommand{\sps}{\(S_{PS}\)}

\title[Misalignments Between Peer Supporters and Experts in LLM-Supported Interactions]{``Is This Really a Human Peer Supporter?": Misalignments Between Peer Supporters and Experts in LLM-Supported Interactions}

\author{Kellie Yu Hui Sim}
\email{kellie_sim@mymail.sutd.edu.sg}
\orcid{0009-0005-6451-7089}
\affiliation{
  \institution{Singapore University of Technology and Design}
  \country{Singapore}
}
\authornote{Corresponding author. This is a preprint of the paper accepted at CSCW 2026. The final version will be available in the ACM Digital Library.}

\author{Roy Ka-Wei Lee}
\email{roy_lee@sutd.edu.sg}
\orcid{0000-0002-1986-7750}
\affiliation{
  \institution{Singapore University of Technology and Design}
  \country{Singapore}
}

\author{Kenny Tsu Wei Choo}
\email{kenny_choo@sutd.edu.sg}
\orcid{0000-0003-3845-9143}
\affiliation{
  \institution{Singapore University of Technology and Design}
  \country{Singapore}
}

\renewcommand{\shortauthors}{Sim et al.}

\begin{abstract}
  Mental health is a growing global concern, prompting interest in AI-driven solutions to expand access to psychosocial support.
\emph{Peer support}, grounded in lived experience, offers a valuable complement to professional care.
However, variability in training, effectiveness, and definitions raises concerns about quality, consistency, and safety.
Large Language Models (LLMs) present new opportunities to enhance peer support interactions, particularly in real-time, text-based interactions. 
We present and evaluate an AI-supported system with an LLM-simulated distressed client (\client{}), context-sensitive LLM-generated suggestions (\suggestions{}), and real-time emotion visualisations.
2 mixed-methods studies with 12 peer supporters and 6 mental health professionals (i.e., experts) examined the system’s effectiveness and implications for practice.
Both groups recognised its potential to enhance training and improve interaction quality.
However, we found a key tension emerged: while peer supporters engaged meaningfully, experts consistently flagged critical issues in peer supporter responses, such as missed distress cues and premature advice-giving. 
This misalignment highlights potential limitations in current peer support training, especially in emotionally charged contexts where safety and fidelity to best practices are essential.
Our findings underscore the need for standardised, psychologically grounded training, especially as peer support scales globally.
They also demonstrate how LLM-supported systems can scaffold this development--if designed with care and guided by expert oversight. 
This work contributes to emerging conversations on responsible AI integration in mental health and the evolving role of LLMs in augmenting peer-delivered care.
\end{abstract}




\maketitle

\section{Introduction}
Peer support has become an increasingly prominent form of mental health care, particularly within community organisations, educational institutions, and online platforms. It is often delivered by volunteers or paraprofessionals—individuals with structured training and lived experience who provide non-clinical emotional support without professional licensure. Peer support emphasises emotional validation, solidarity, and mutual understanding rather than diagnosis or treatment~\cite{meadPeerSupportWhat2004, repperReviewLiteraturePeer2011}. However, despite its growing adoption, peer support remains characterised by substantial variation in training, role expectations, and interactional norms across organisations and contexts. Unlike professional mental health care, where standards of practice and accountability are relatively well defined, peer support often relies on implicit, situational judgements about what constitutes appropriate or effective support. This ambiguity can enable flexibility and authenticity, but also introduces tension in emotionally charged interactions where safety, trust, and perceived quality of support are at stake.

These tensions are particularly salient given the close proximity between peer support and professional mental health practices. Clinicians are trained to assess risk, regulate emotional arousal, and guide conversations toward structured therapeutic goals, whereas peer supporters are often encouraged to "hold space", share selectively from personal experience, and avoid prescriptive advice~\cite{meadPeerSupportWhat2004}. In practice, however, the boundaries between peer and professional approaches are increasingly blurred, as peer support programmes draw on professionalised frameworks such as Psychological First Aid (PFA)~\cite{shultzPsychologicalFirstAid2014a} and Motivational Interviewing (MI)~\cite{rollnickWhatMotivationalInterviewing1995}, as well as shared digital infrastructures~\cite{leeCounsellorsAlarmedUnqualified2025, leeSafetyMentalHealth2025}. As a result, expectations about what constitutes "good" support may diverge across stakeholders, even within the same interaction.

Recent advances in artificial intelligence (AI), particularly large language models (LLMs), have accelerated the development of tools intended to scaffold mental health conversations, including systems that simulate distressed clients or generate empathetic response suggestions~\cite{hsuHelpingHelperSupporting2025, liuComPeerGenerativeConversational2024a, morrisArtificiallyEmpathicConversational2018, pengExploringEffectsTechnological2020}. While often positioned as supportive aids for volunteers or novices, such systems are typically grounded in professional therapeutic norms, such as structured empathy and staged intervention~\cite{liUnderstandingTherapeuticRelationship2024, sharmaFacilitatingEmpathicConversations2021, tananaHowYouFeel2021}. When introduced into peer support contexts, these assumptions may not simply assist supporters, but also surface underlying differences in how peer and professional care are enacted and evaluated.

In this paper, we argue that AI-mediated peer support interactions offer an analytic lens for examining how differences between peer supporters’ practices and professional mental health standards become visible in practice. Rather than treating AI assistance as a training or performance-enhancing tool, we examine how its presence reshapes interactional dynamics, cognitive and emotional labour, and judgements of response quality across stakeholder groups. Specifically, we ask: \textbf{(RQ1)} How do peer supporters engage with an LLM-simulated client, LLM-generated suggestions, and emotional state visualisations during multi-turn, real-time peer support interactions, and what shapes their perceptions of these components’ usefulness? And \textbf{(RQ2)} How do mental health professionals interpret and assess peer supporter responses and the LLM-supported components (see RQ1), and where do their assessments converge or diverge from peer supporters’ perspectives?

To address these questions, we conducted two mixed-methods studies using an AI-supported peer support system comprising three components: (1) a large language model (LLM)-simulated distressed client (\client), (2) real-time, categorised response suggestions (\suggestions{}), and (3) emotional state visualisations derived from the client’s messages. In Study~1, twelve trained peer supporters engaged in live, text-based conversations with \client while interacting with these AI features. In Study~2, six mental health professionals reviewed recordings of these sessions and reflected on the peer supporters’ responses and the role of the AI-supported components.

Across both studies, we find that AI-mediated interactions surface patterned areas of both convergence and divergence between peer supporters and mental health professionals, particularly around safety, timing, authenticity, and the perceived appropriateness of AI-supported responses. Building on these findings, this paper makes the following contributions:
\begin{itemize}
    \item We provide empirical evidence of how AI-mediated peer support interactions surface both shared ground and misalignments between peer supporters’ practices and mental health professionals' perspectives.
    \item We offer a detailed account of how AI-generated suggestions and emotional state visualisations reshape peer supporters’ cognitive and emotional labour during real-time support conversations.
    \item We derive design implications for CSCW systems that mediate mental health support across roles with differing expertise and accountability, highlighting the risks of uncritically embedding professionalised assumptions into peer support technologies.
\end{itemize}
\section{Related Work}
\subsection{Peer Support as a Distinct Interactional Practice}
Peer support refers to the mutual provision of emotional or practical assistance between individuals with shared lived experiences of mental health challenges~\cite{meadPeerSupportWhat2004, repperReviewLiteraturePeer2011}.
Unlike professional mental health services--such as psychotherapy or crisis intervention--peer support is not oriented toward diagnosis or treatment, but toward empathy, relatability, and shared understanding~\cite{solomonPeerSupportPeer2004}.

Although peer support may draw on practices familiar to professional care, it is grounded in commitments to mutuality, non-hierarchical exchange, and the legitimacy of lived experience as a form of expertise~\cite{leePeerSupportMental2019}.
These values shape how support is enacted in practice and how interactional choices are interpreted, often in ways that differ from professional mental health contexts.
Rather than adhering to codified protocols, peer support interactions are frequently negotiated in situ, with judgements about appropriateness emerging from the relational dynamics of the exchange~\cite{kirkegaardExperientialKnowledgeMental2022, scottHavingThoseConversations2011, simSaidThingsNeeded2025}.

In practice, peer support spans a wide range of organisational forms, from informal volunteer-led communities to more formalised roles embedded within public mental health systems~\cite{leePeerSupportMental2019, frankeImplementingMentalHealth2010, shalabyPeerSupportMental2020}.
Across many contexts--particularly in Asia and the Global South--peer support often remains semi-formal or community-based, shaped by local norms, stigma, and social hierarchies~\cite{chuaYoungAdultsMental2024, greenInfluenceCulturalNorms2024, kaurExploringPotentialPeer2025}. 
As a result, what constitutes \dquote{appropriate} or \dquote{effective} support may vary substantially across settings and stakeholders.

These characteristics have important implications for AI-supported peer support.
Many AI systems implicitly encode professional assumptions about emotional expression, pacing, and safety.
When introduced into peer support settings, such systems may surface tensions between professionalised evaluative standards and peer supporters' situated judgements.
In this paper, we treat peer support as a distinct interactional practice rather than a simplified analogue of professional care, enabling us to examine how AI-mediated interactions reveal misalignments between peer supporters and mental health professionals.

\subsection{From Online Peer Support to AI-Mediated Peer Support}
As peer support has increasingly shifted online, many practices traditionally grounded in face-to-face interaction have been reconfigured through text-based, asynchronous, and anonymous communication~\cite{raylandSocialNetworkPeer2023}.
Digital platforms such as \textit{7 Cups}\footnote{\url{https://www.7cups.com}},
\textit{TalkLife}\footnote{\url{https://www.talklife.com}}, and national crisis
services (e.g., \textit{Samaritans}\footnote{\url{https://www.samaritans.org/}},
\textit{Lifeline}\footnote{\url{https://www.lifeline.org.au/}}, and the
\textit{988 Suicide and Crisis Lifeline}\footnote{\url{https://988lifeline.org/}})
connect volunteer supporters with help-seekers through structured conversations.
Similar ecosystems have emerged in Singapore, spanning moderated forums (\textit{let's talk}\footnote{\url{https://letstalk.mindline.sg}}), text-based crisis intervention (\textit{SOS CareText}\footnote{ \url{https://www.sos.org.sg/pressroom/sos-launches-new-text-based-service-for-those-in-distress/}}), and blended volunteer--professional models like \textit{SAFEHOUSE}\footnote{\url{https://www.limitless.sg/safehouse}}.
While these platforms broaden access, they also foreground interactional challenges central to peer support, including how empathy, responsiveness, and authenticity are enacted in sustained, text-based exchanges.

Beyond formal services, informal peer support communities have flourished on platforms such as \textit{Reddit}, \textit{Discord}, \textit{Facebook}, \textit{WhatsApp}, and \textit{Telegram}~\cite{kimSupportersFirstUnderstanding2023, gauthierWillNotDrink2022, perepezkoInsteadYoureGoing2024, gohYoungPeopleCite2023, yeshua-katzRoleCommunicationAffordances2021, lambton-howardBlendingEverydayLife2021, yeoDigitalPeerSupport2023, rolandoTelegramSpacePeerLed2023}.
These spaces are typically self-organised, minimally moderated, and valued for anonymity, flexibility, and reduced stigma~\cite{kruzanInvestigatingSelfinjurySupport2021, iftikharTogetherNotTogether2023}.
Importantly, they function as sites where norms of peer support are negotiated in practice rather than enforced through formal protocols, often serving as initial touchpoints when professional services are inaccessible or stigmatised~\cite{naslundFutureMentalHealth2016}.
Prior work suggests such interactions may support psychological well-being, though outcomes vary widely across contexts~\cite{yeoDigitalPeerSupport2023}.

At the same time, online peer support exposes tensions inherent to text-mediated interaction. The absence of non-verbal cues constrains how distress is interpreted and how empathy is conveyed~\cite{iftikharTogetherNotTogether2023, terryEmergingIssueDigital2016}.
Volunteer supporters frequently face emotional strain and burnout, while training and supervisory structures remain uneven~\cite{chenScaffoldingOnlinePeersupport2021}.
Moreover, interactional techniques commonly promoted in training--such as reflective listening or paraphrasing--may be experienced as inauthentic or counterproductive if applied rigidly or without sensitivity to the relational context~\cite{goldbergPsychotherapistsImproveTime2016}.
These challenges highlight that ``effective'' support in peer contexts cannot be reduced to the correct application of techniques alone.

In response, researchers have begun exploring AI systems to scaffold peer support interactions, for example by offering real-time guidance, message enhancements, or context-aware suggestions~\cite{liuComPeerGenerativeConversational2024a, pengExploringEffectsTechnological2020, morrisArtificiallyEmpathicConversational2018, youngRoleAIPeer2024}.
Many such systems draw on counselling frameworks--including motivational interviewing, CBT, and active listening--to model supportive responses~\cite{sharmaFacilitatingEmpathicConversations2021, tananaHowYouFeel2021}, while systems such as \textit{CARE} integrate conversation history and emotion recognition to tailor guidance~\cite{hsuHelpingHelperSupporting2025}.
Related work has also examined therapeutic alliance and relational alignment in AI-supported counselling, emphasising affective bonding and collaborative engagement~\cite{liUnderstandingTherapeuticRelationship2024}.

However, much of this work implicitly evaluates AI assistance against professional or instructional standards, often within controlled or scripted settings~\cite{hsuHelpingHelperSupporting2025, sharmaHumanAICollaboration2023, liUnderstandingTherapeuticRelationship2024}. 
As a result, less is known about how AI-generated guidance is interpreted and negotiated within the situated realities of volunteer peer support, where norms of appropriateness, authenticity, and care are fluid and contested.
In particular, prior research has paid limited attention to how peer supporters themselves engage, adapt, or resist AI suggestions during live, multi-turn interactions, or how these practices align--or conflict--with professional expectations.

Building on this gap, our study examines how trained peer supporters engage with LLM-generated suggestions during real-time, multi-turn conversations with an LLM-simulated distressed client.
By additionally incorporating interviews with mental health professionals, we surface how peer support practices and AI-mediated interactions are evaluated against professional standards, enabling a comparative examination of normative misalignments that are often assumed away in prior AI-assisted peer support research.

\subsection{LLM-Driven Conversational Agents and Simulated Clients in Mental Health}
Conversational agents (CAs) have become common in digital mental health interventions~\cite{liSystematicReviewMetaanalysis2023}, spanning cognitive behavioural therapy (CBT)-based chatbots~\cite{fitzpatrickDeliveringCognitiveBehavior2017} and expressive writing tools~\cite{parkWroteIfWere2021} to solution-focused counselling approaches~\cite{fulmerUsingPsychologicalArtificial2018}, and complement developments of broader digital mental health technologies like virtual reality, mobile platforms, and self-guided tools~\cite{hollisAnnualResearchReview2017, chenPromotingMentalHealth2024, emmelkampVirtualRealityTherapy2020, limdingfengSocialBehaviourUnderstanding2020}.
Real-world evaluations, such as of \textit{Wysa}, suggest CAs can alleviate depressive symptoms through empathetic text-based interaction~\cite{inksterEmpathyDrivenConversationalArtificial2018}.

LLMs have expanded CA capabilities, enabling more fluid and context-aware dialogue than scripted systems~\cite{kovacevicPersonalityDimensionsGPT32024}.
Systems like \textit{MindfulDiary} use journaling for therapeutic reflection~\cite{kimMindfulDiaryHarnessingLarge2024}, while \textit{ComPeer} integrates peer support strategies and conversation history to tailor responses to emotional needs~\cite{liuComPeerGenerativeConversational2024a}.
By emulating principles such as empathy, emotional validation, reflective listening, and motivational interviewing~\cite{sharmaFacilitatingEmpathicConversations2021, chaszczewiczMultiLevelFeedbackGeneration2024}, LLM-based CAs approximate trained supporter responses, potentially lowering barriers to care.
At the same time, individuals increasingly turn to general-purpose LLMs like ChatGPT for emotional support, drawn by immediacy and anonymity~\cite{alavanzaSingaporeYouthTurn2025}, though such unsupervised uses raise safety concerns~\cite{louChatbotsSeeGreater2023, martinengoEvaluationChatbotdeliveredInterventions2022}.

Despite their promise, LLMs face limitations in handling ambiguity, emotional pacing, and high-stakes disclosures, creating risks of over-reliance or misplaced trust~\cite{songTypingCureExperiences2024, tananaHowYouFeel2021}.
Ethical and effective use requires careful design and, often, human oversight.

In parallel, LLMs have been used to simulate \emph{patients} or \emph{clients} for training and evaluation.
These systems aim to replicate the conversational behaviours, emotional trajectories, and presentation styles of real clients, creating safe and scalable environments for therapists, clinical students, and researchers to practise and refine their skills~\cite{steenstraScaffoldingEmpathyTraining2025, yangConsistentClientSimulation2025, wangClientCenteredAssessmentLLM2024, wangPATIENTPsUsingLarge2024}.
Such simulations typically maintain consistent personas across multi-turn conversations, enabling standardised practice with clients across diverse levels of psychological distress~\cite{wangClientCenteredAssessmentLLM2024} that follow structured protocols with embedded cues for active listening, de-escalation, and rapport-building.
LLMs have also been used to simulate \emph{therapists} or \emph{counsellors}~\cite{kampmanConversationalSelfPlayDiscovering2025, yangConsistentClientSimulation2025, chiuComputationalFrameworkBehavioral2024}, though this work largely targets professional contexts.

Much less is known about using simulated clients in peer support, where conversations are often less structured, highly emotional, and facilitated by volunteers with minimal training.
These settings introduce unique design challenges for simulated clients, demanding calibrated emotional pacing, conversational realism, and interpersonal tone to align with non-professionals’ capacities and expectations.

Our work extends existing approaches into the peer support domain, where the stakes, expectations, and relational dynamics differ substantially. 
We capture and analyse real-time, multi-turn interactions across peer supporters and a LLM-simulated distressed client with visualised emotional states, alongside live LLM-generated suggestions intended to scaffold their responses in real time.
Expert evaluations guide further analysis of these components, as well as the relational sensitivity shown by peer supporters.
This dual lens enables a more holistic understanding of how simulated clients may support training, reflection, and future system design within peer support settings.
\section{Overview}~\label{section:overview}
\subsection{System Overview and Study Design}
\begin{figure}[htbp]
    \centering
    \includegraphics[width=0.95\linewidth]{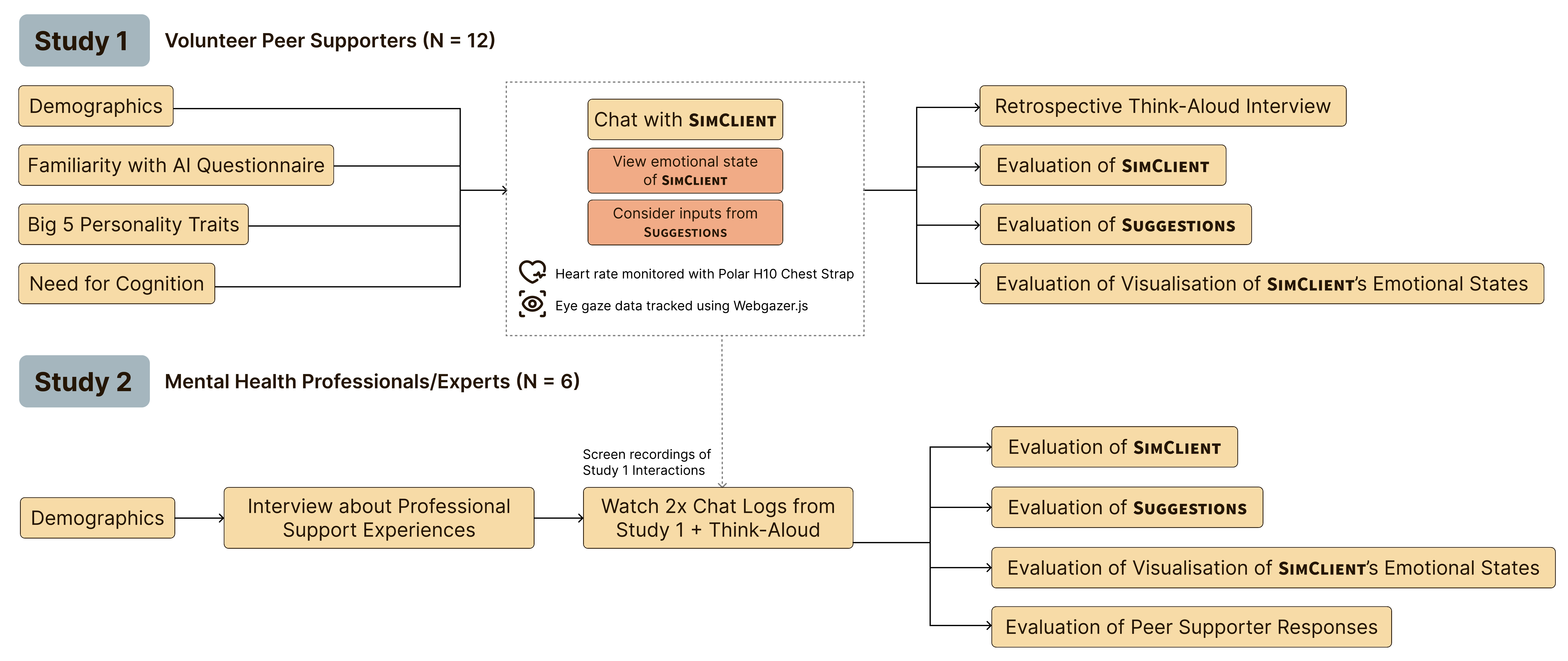}
    \caption{Overview of the studies conducted.
    Study 1 included 12 peer supporters engaging in real-time chats with \client{}, using \suggestions{} and emotional state visualisations, accompanied by retrospective think-aloud interviews and physiological/behavioural monitoring. 
    Study 2 involved 6 experts (clinical psychologist, counsellors, medical social worker) evaluating Study 1 logs and reflecting on \client{}, \suggestions{}, and emotional state visualisations.}
    \Description{Diagram showing the workflow of the studies conducted.
    Study 1 (N=12 peer supporters): demographics, chat with a simulated client, \client{}, while using AI-generated suggestions (\suggestions{}) and visualised emotional states, retrospective think-aloud interviews, and physiological measures (eye-tracking and heart rate monitoring).
    Study 2 (N=6 expert professionals): demographics, interviews on professional support experiences, and review of Study 1 logs with evaluations of \client{}, AI suggestions, and visualisations.}
    \label{fig:study-overview}
\end{figure}

To understand how peer supporters engage with AI-mediated support during live interactions, we developed and evaluated an LLM-supported system comprising three components inspired by prior work: an LLM-simulated distressed client (\client{}), context-sensitive LLM-generated suggestions (\suggestions{}), and real-time emotional state visualisations.
These were deployed in a text-based chat interface to support psychologically realistic peer support scenarios.
To investigate the system’s effectiveness and implications for peer support practice, we conducted two studies (see Figure~\ref{fig:study-overview}):
\begin{enumerate}
    \item \textbf{Study 1: Lab Study with Peer Supporters.}  
    12 trained peer supporters engaged in synchronous, multi-turn conversations with \client{}, supported by real-time emotional state visualisations and \suggestions{}.
    \item \textbf{Study 2: Expert Evaluation of Study 1's Interactions.}  
    6 mental health professionals reviewed anonymised recordings from Study 1 and evaluated \client{}, the quality of \suggestions{}, and the utility of emotional state visualisations. They also raised considerations for future deployments.
\end{enumerate}

Together, these studies examined how AI tools shape real-time peer support while preserving peer supporters’ central role, aligned with calls for multi-layered, context-sensitive evaluations of AI in mental health care~\cite{simEnvisioningAIEnhancedMental2025}.
Both studies received Institutional Review Board approval (SUTD IRB-24-00670) and adhered to ethical research guidelines.
Participants were compensated approximately USD\$15 (Study 1) and USD\$23 (Study 2).

\begin{figure}[htbp]
    \centering
    \includegraphics[width=\textwidth]{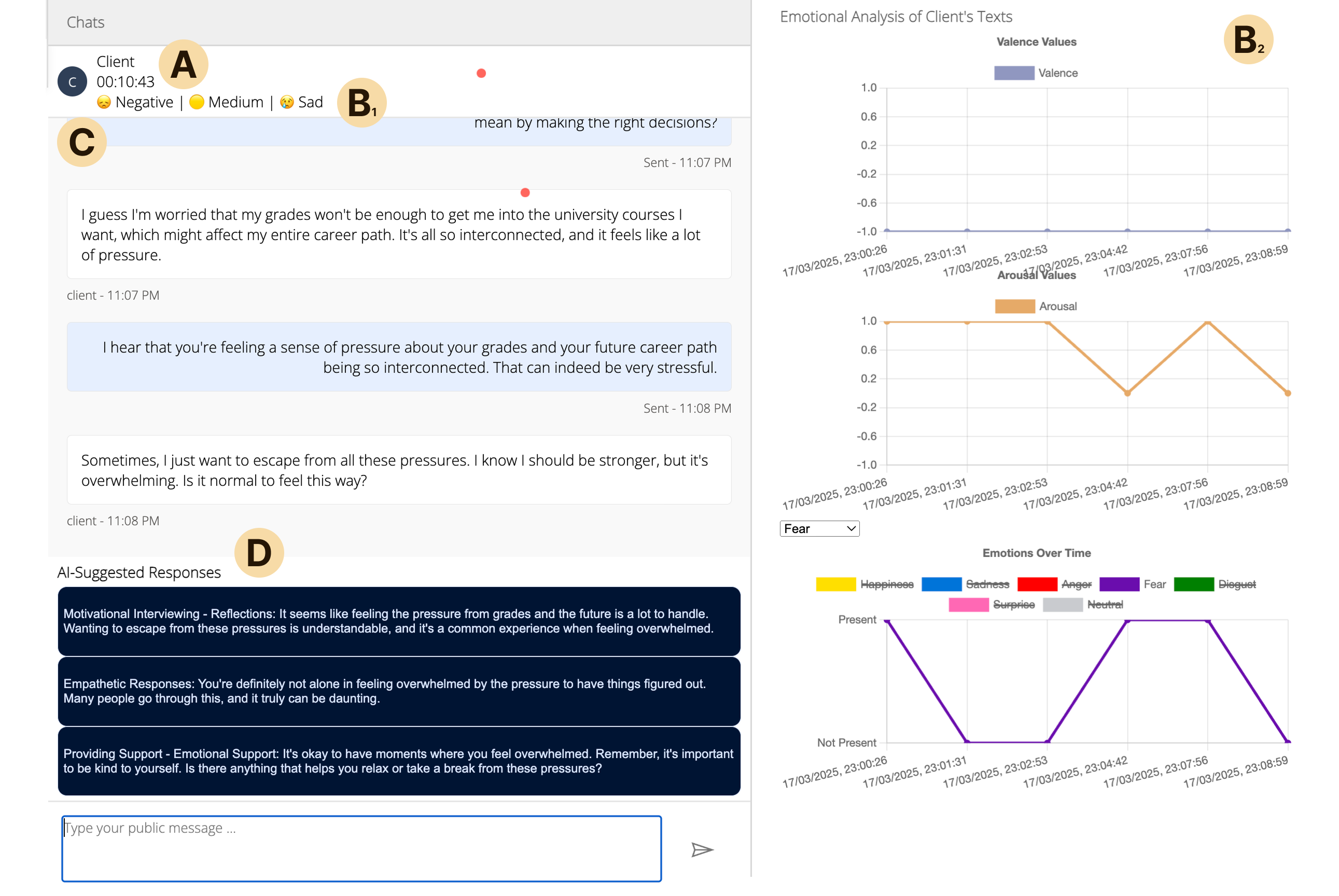}
    \caption{Chat interface used in the study. Participants interact with \client{} by composing text-based responses to client messages. The interface provides real-time emotional feedback, including (A) the client's label and chat duration, (B\textsubscript{1}) an emoji and label-based summary of emotional states, and (B\textsubscript{2}) time-series graphs of valence, arousal, and categorical emotions. The main chat window (C) shows the ongoing exchange. Below, \suggestions{} are displayed in three categories for participants' choosing: motivational interviewing (\smi{}), empathetic responses (\ser{}), and emotional support (\sps{}).}
    \Description{Screenshot of a chat interface used in a study where participants respond to an LLM-simulated client. (A) shows the label "Client" with a chat timer (e.g., "00:10:43"). (B\_1) presents an emoji-based summary of the client's emotional state (e.g., Negative, Medium, Sad), and (B\_2) contains time-series graphs displaying emotional trends, such as valence, arousal, and emotion presence. The central chat panel (C) contains dialogue where the client expresses academic stress, and participants are expected to respond. Below the chat, section (D) provides LLM-generated suggestions (\suggestions{}) that participants could choose to use in three categories: motivational interviewing (\smi{}), empathetic responses (\ser{}), and emotional support (\sps{}), to guide participants in composing replies.}
    \label{fig:chat-interface}
\end{figure}

Figure~\ref{fig:chat-interface} illustrates the chat interface developed for Study 1 and evaluated in Study 2.
It comprised: (A) a client label and chat timer, (B\textsubscript{1}) an emoji and summary of \client{}’s emotional state, (B\textsubscript{2}) time-series graphs of dimensional and categorical emotions, (C) the chat window showing the exchange, and (D) \suggestions{} displayed in three categories (\smi{}, \ser{}, \sps{}).

\subsection{Design of LLM-Simulated Client: \client{}}
\client{} was introduced to enable controlled, repeatable peer support scenarios while preserving affective realism.
Prior work shows that conversational simulations can support the development of empathic and relational skills in emotionally charged contexts~\cite{steenstraScaffoldingEmpathyTraining2025, yangConsistentClientSimulation2025, wangClientCenteredAssessmentLLM2024, wangPATIENTPsUsingLarge2024}.
Simulations provide a low-risk environment for practising turn-taking, reflective listening, and empathetic questioning.

\client{} was powered by GPT-4 Assistants (gpt-4o-2024-08-06) and designed to emulate an emotionally expressive support-seeker that engaged in a multi-turn conversational exchange with participants. 
Responses were generated dynamically throughout the conversation, guided by structured conversational phases (see Appendix~\ref{appendix:phase-2-simclient-suggestions-instructions}). 

\subsection{Design of \client{}'s Emotional State Visualisations}


Each \client{} message was paired with a computed emotional state derived from two complementary models~\cite{simUnderstandingEmotionsEngaged2024} grounded in established affective theories:
\begin{itemize}
    \item \textbf{Dimensional}: valence (-1 to +1) and arousal (-1 to +1), following the circumplex model of affect~\cite{posnerCircumplexModelAffect2005}, capturing affective tone and intensity that was visualised using sentiment labels and time-series graphs (Figure~\ref{fig:chat-interface}, B\textsubscript{1}, B\textsubscript{2}).
    \item \textbf{Categorical}: one discrete emotion from the following: \textit{sadness}, \textit{fear}, \textit{anger}, \textit{disgust}, \textit{happiness}, \textit{surprise} (adapted from \citet{ekman-1999-emotions}), and neutral, also visualised (Figure~\ref{fig:chat-interface}, B\textsubscript{1}, B\textsubscript{2}),
\end{itemize}

These were computed from the emotional intent embedded in each GPT-generated message (see Appendix~\ref{appendix:phase-2-simclient-suggestions-instructions}).
Visualisations were designed to support interpretation of affective shifts in text-based peer support, where paralinguistic cues are absent, leading to increased risk of misinterpreting distress or emotional nuance.
Combining dimensional trajectories with categorical labels enabled \client{} to exhibit dynamic emotional patterns.
For example, higher-arousal negative states were associated with more urgent or fragmented message styles.
This coupling aimed to support supporters in monitoring affective change, calibrating pacing, and recognising moments requiring heightened attunement.

\subsection{Design of LLM-Generated Suggestions: \suggestions{}}
\suggestions{} was designed to scaffold peer supporters’ cognitive and linguistic work during emotionally demanding interactions.
Prior research suggests that real-time scaffolding can reduce cognitive load and support conversational alignment in supportive dialogue~\cite{hsuHelpingHelperSupporting2025, sharmaFacilitatingEmpathicConversations2021, tananaHowYouFeel2021}.
Such scaffolds are particularly beneficial when users must simultaneously track emotional cues and formulate responses.

\suggestions{} was powered by GPT-4 Assistants (gpt-4o-2024-08-06). 
Three suggestions were generated after each \client{} message and could be \textit{adopted}, \textit{adapted}, or \textit{ignored}.
Table~\ref{tab:suggestion-categories} summarises the categories and example outputs for the message: \dquote{I’ve been feeling really anxious about school lately and unsure about my future. It’s like everyone is moving forward and I’m stuck.}


\begin{table}[htbp]
\centering
\small
\caption{Suggestion Categories and Examples}
\label{tab:suggestion-categories}
\renewcommand{\arraystretch}{1.2}
\begin{tabular}{p{3.3cm} p{4.5cm} p{5cm}}
\toprule
\textbf{Suggestion Category} & \textbf{Intended Function} & \textbf{Example Suggestion} \\
\midrule
Motivational Interviewing (\smi{}) \cite{hsuHelpingHelperSupporting2025, shahModelingMotivationalInterviewing2022}
& Encourage reflection and elaboration through open-ended questioning 
& Can you tell me more about what specifically is making you anxious about school and the future? \\

Empathetic Responses (\ser{}) \cite{sharmaFacilitatingEmpathicConversations2021, majumderMIMEMIMickingEmotions2020}
& Validate emotions and support rapport-building 
& It sounds like you're feeling overwhelmed with the uncertainty and pressure of what's to come. That’s completely understandable. \\

Emotional or Informational Support (\sps{}) \cite{pengExploringEffectsTechnological2020, kruzanInvestigatingSelfinjurySupport2021}
& Provide reassurance, normalisation, or relevant guidance 
& You’re not alone in these feelings. Let’s work through them together. \\
\bottomrule
\end{tabular}
\end{table}

Further examples of prompt structure and system output are provided in Appendices~\ref{appendix:phase-2-simclient-suggestions-instructions} and~\ref{appendix:phase-2-case-studies}.
\section{Study 1: Lab Study with Peer Supporters}

\subsection{Methodology}
\subsubsection{Participants}
We recruited 12 participants (5 female), all aged 18 and above ($mean = 34.04$, $SD = 14.67$), each with prior experience and training in peer support and/or PFA (see Table~\ref{tab:study-1-demographics}).
Participants were drawn from both community- and institutional-based peer support programmes across Singapore, with experience spanning online ($n = 5$), offline ($n = 2$), and hybrid ($n = 5$) modalities.
Most held at least a Bachelor's degree ($n = 11$), including two with Psychology degrees.
Peer support experience ranged from 1 to over 10 years ($mean = 3.6$ years across modalities).
\begin{table*}[htbp]
\centering
\small
\caption{Demographic and Peer Support Experience Details}
    \label{tab:study-1-demographics}
    \begin{tabular}{@{}p{0.5cm}p{0.8cm}p{0.9cm}p{1.35cm}p{1.3cm}p{2cm}p{1.9cm}p{2.1cm}@{}}
    \toprule
    \textbf{ID} & \textbf{Age} & \textbf{Gender} & \textbf{Education} & \textbf{Ethnicity} & \textbf{Employment} & \textbf{Psychology Background?} & \textbf{Experience (Type and \newline Duration)} \\
    \midrule
    \duskytide{} & 25-34 & Female & Bachelor's & Chinese & Self-employed & No & Online (\textasciitilde6 years) \\
    \zoomplane{}{} & 18-24 & Female & Bachelor's & Indian & Student & Yes & Both (\textasciitilde4 years offline, 1 year online) \\    
    \windsong{} & 25-34 & Male & Diploma & Chinese & Part-time & No & Both (\textasciitilde1-2 years) \\
    \amberlight{} & 65-74 & Male & Master's & Chinese & Retired & No & Offline (\textasciitilde6.5 years) \\
    \brightstar{} & 25-34 & Male & Bachelor's & Chinese & Full-time & No & Both (\textasciitilde2-3 years) \\
    \butterflywave{} & 25-34 & Male & Bachelor's & Chinese & Full-time & No & Offline (\textasciitilde1 year) \\
    \bluemoon{} & 45-54 & Female & Bachelor's & Chinese & Full-time & No & Both (\textasciitilde5-10 years) \\
    \greenship{} & 18-24 & Female & Bachelor's & Chinese & Student & No & Online (\textasciitilde1 year) \\
    \ironpetal{} & 25-34 & Male & Master's & Indian & Full-time & No & Both (\textasciitilde3.5 years) \\
    \earnprom{} & 25-34 & Female & Bachelor's & Chinese & Full-time & Yes & Online (\textasciitilde1 year) \\
    \redballoon{} & 45-54 & Male & Bachelor's & Chinese & Self-employed & Yes & Online (\textasciitilde1 year) \\
    \greenleaf{} & 18-24 & Female & Bachelor's & Indian & Full-time & Yes & Online (\textasciitilde3-4 years) \\
    \bottomrule
    \end{tabular}
\end{table*}


\subsubsection{Study Procedure}
Each session comprised three segments: Pre-Chat, Chat, and Post-Chat, all conducted in person and screen-and audio-recorded.

\paragraph{\textbf{Pre-Chat Segment.}}
Participants completed a demographics questionnaire, personality scales (BFI-2-XS~\cite{sotoShortExtrashortForms2017, mccraeChapter31Conceptions1997, mccraeIntroductionFivefactorModel1992}), Need for Cognition-s~\cite{cacioppoEfficientAssessmentNeed1984, cacioppoNeedCognition1982}, and a questionnaire assessing familiarity with generative AI tools such as ChatGPT.

\paragraph{\textbf{Chat Segment.}}
Participants engaged in a multi-turn text-based conversation with \client{}, responding freely or drawing from three categories of LLM-generated suggestions (\smi{}, \ser{}, \sps{}).
Physiological and attentional (eye-tracking) data were collected using a Polar H10 chest strap (HRV) and WebGazer.js~\cite{papoutsaki2016webgazer}.
These measures were collected on an exploratory basis as potential indicators of participant stress and engagement.
All chat transcripts and recordings of the interaction were retained for subsequent expert evaluation in Study 2.

\paragraph{\textbf{Post-Chat Segment.}}
Participants took part in retrospective think-aloud interviews while reviewing their session recordings, guided by facilitator-noted moments on decision-making, suggestions usage, and perceived challenges.
They rated \client{}’s responses on naturalness in terms of fluency and ease (yes/no), human-likeness (non-robotic qualities), and realism (resemblance to real-world clients) on 7-point Likert scales, and reflected on the usefulness of \suggestions{} and emotional state visualisations.
These measures assessed \client{}'s ability to simulate a believable peer support interaction and participants' general impressions of its responses.

\subsubsection{Data Analysis}
Interview transcripts were analysed by one researcher using an iterative, reflexive thematic analysis process involving repeated close readings, memo writing, and progressive refinement of codes and themes.
Inter-rater reliability was not calculated, as it is not appropriate for interpretivist analyses of this kind~\cite{mcdonaldReliabilityInterraterReliability2019}.
Analytical rigour was supported through reflexive documentation and transparency of analytic decisions.

Physiological and behavioural data (HRV and eye-tracking) were analysed on an exploratory basis. 
HRV features (RMSSD, SDNN, LF/HF ratio) were computed using Kubios HRV Standard following artefact correction and segmentation.
Eye-tracking analyses examined gaze fixation duration and attentional engagement during the chat.
\subsection{Results}
\subsubsection{Engagement with \client{}}
\paragraph{\textbf{Message Length and Conversational Pacing.}}
Participants and \client{} engaged in structured, turn-based exchanges, averaging 34 messages over approximately 30 minutes (Table~\ref{tab:study-1-session-stats}).
While conversation length and participant verbosity varied substantially across sessions, \client{} maintained relatively consistent response timing and message length.

\begin{table}[htbp]
    \centering
    \small
    \caption{Session Message Counts and Durations for Study 1 (N = 12)}
    \Description{Table summarising the message exchange and duration statistics across 12 peer support sessions. On average, participants and \client{} exchanged 33.5 messages over 29 minutes and 43 seconds. The number of messages varied widely (SD = 16.09), reflecting individual differences in pacing and response habits. Session lengths also varied (SD = 11 minutes 58 seconds), with some sessions lasting as little as 11 minutes and others exceeding 50 minutes.} 
    \begin{tabular}{lcccccc}
    \toprule
    \textbf{Session Breakdown} & \textbf{Mean} & \textbf{SD} & \textbf{Min} & \textbf{Max} & \textbf{Median} & \textbf{IQR} \\
    \midrule
    Number of Messages (2-way) & 33.5 & 16.09 & 20 & 74 & 28 & 12 \\
    Length of Conversation (hh:mm:ss) & 00:29:43 & 00:11:58 & 00:11:24 & 00:51:16 & 00:27:32 & 00:12:57 \\
    \bottomrule
    \end{tabular}
    \label{tab:study-1-session-stats}
\end{table}

\begin{figure}[htbp]
    \centering
    \includegraphics[width=\textwidth]{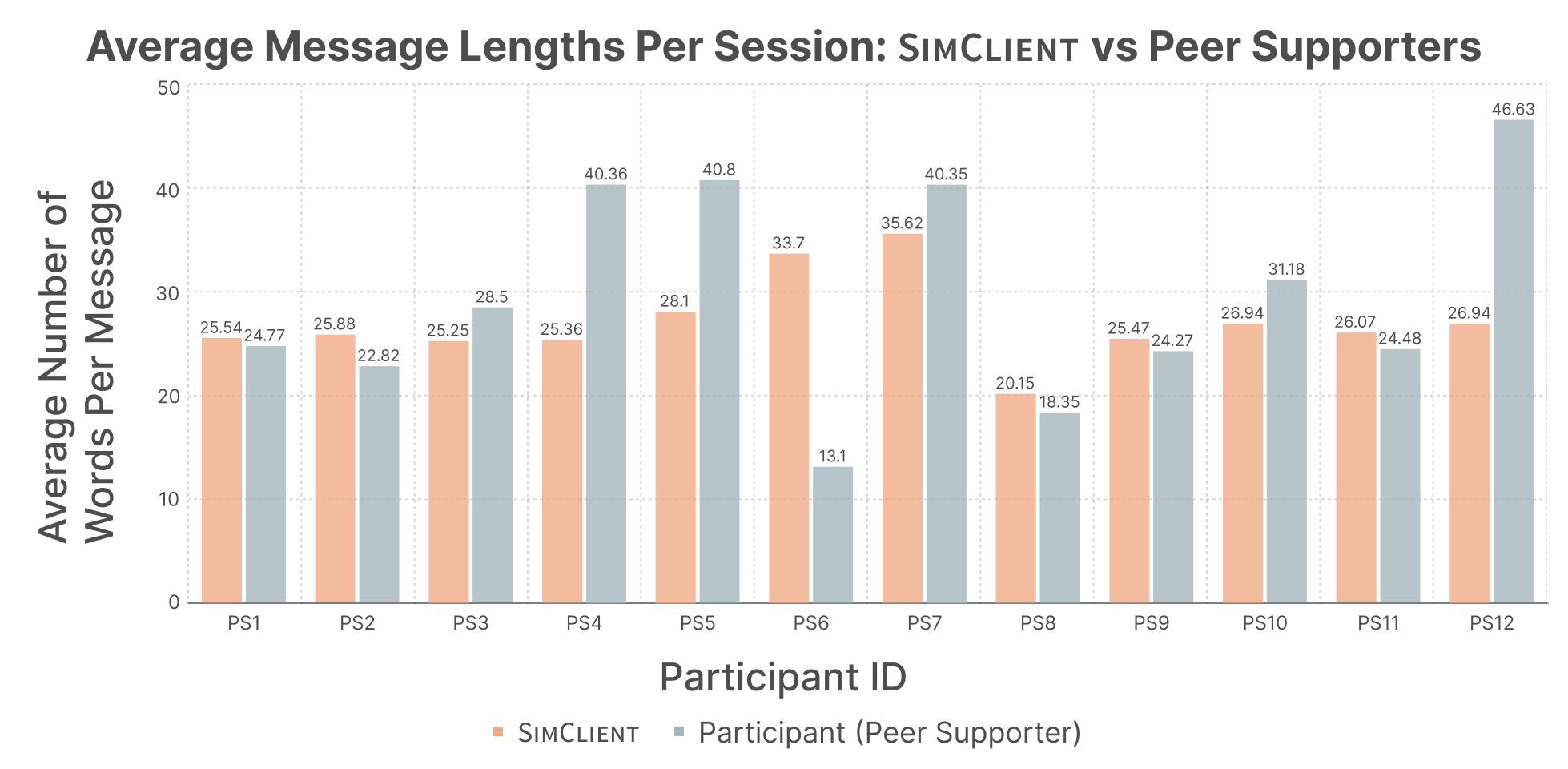}
    \caption{Average message lengths (words per message) by participant in Study~1, showing greater variability among peer supporters compared to the relatively stable responses of \client{}.}
    \Description{Bar chart showing the average number of words per message by all 12 participants in Study 1), comparing peer supporters and \client{}. Each participant is represented on the x-axis. The y-axis shows message length in words. Peer supporter message lengths vary widely across sessions, with some exceeding 40 words per message, while \client{}'s message lengths remain relatively consistent across participants, mostly between 20 and 35 words.}
    \label{fig:phase2-average-message-length}
\end{figure}

At the message level, \client{}’s responses (27.89 words per message (wpm)) were more uniform than participants’ (29.51 wpm), and participant verbosity did not reliably elicit longer replies.
Correlations between participant and \client{} message lengths were weak ($r = 0.18$), and only 3 sessions showed moderate turn-level synchrony ($r \geq 0.5$, $p < 0.05$), indicating limited reciprocal adjustment in pacing (Figure~\ref{fig:phase2-average-message-length}).

Taken together, these findings suggest that \client{} sustained a stable interactional rhythm across sessions, with limited adaptation to individual peer supporter styles.

\paragraph{\textbf{Emergent Strategies and Moments of Hesitation.}}
Across sessions, participants employed peer support and PFA principles--including affirmation, active listening, emotional regulation, and respecting autonomy--while adapting flexibly to perceived cues from \client{}.
Rather than following rigid scripts, participants treated the interaction as a relational process, calibrating tone, pacing, and involvement in response to the unfolding conversation.
Participants avoided advice-giving, and as \zoomplane{} explained (similarly reflected by \windsong{} and \butterflywave{}), their responses were a \dquote{very subconscious effort} to help \client{} feel \dquote{like they are in it together}.

Moments of hesitation revealed \emph{thoughtful restraint}.
Participants paused or withheld follow-up questions when responses felt emotionally vague, signalling attunement to boundaries.
For example, \brightstar{} described choosing not to probe further after a non-committal reply: \dquote{I didn't want to probe again}.
Screen recordings and eye-tracking data corroborated these accounts, showing brief pauses before replies to emotionally vague content.
Such hesitation functioned as an interactional strategy aligned with real-world peer support practices.

\begin{figure}[htbp]
    \centering
    \includegraphics[width=0.9\textwidth]{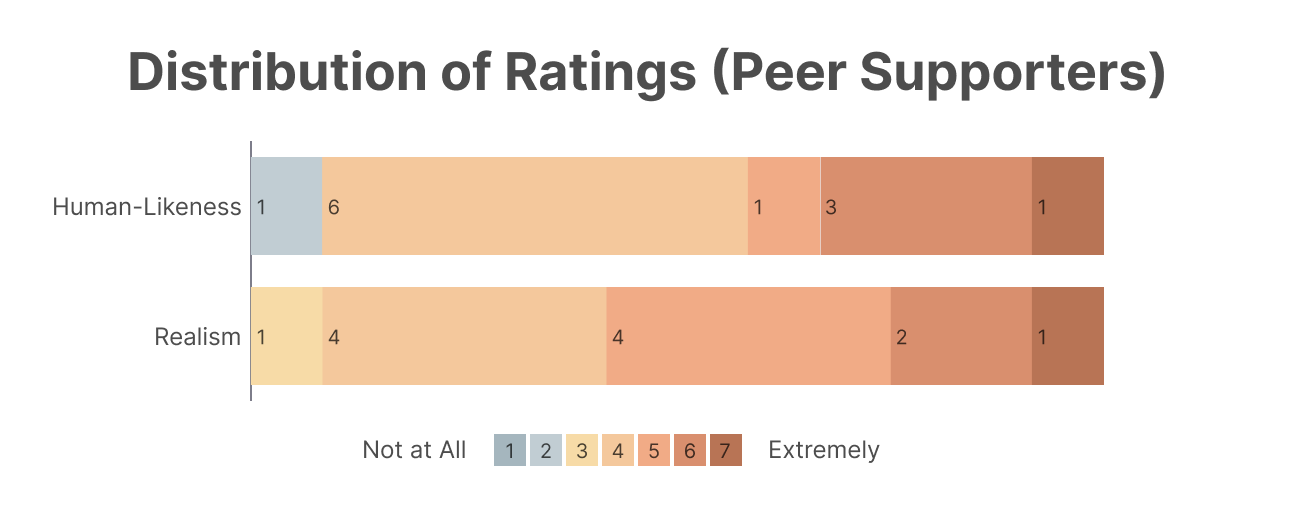}
    \caption{Distribution of peer supporter ratings of \client{} in Study~1 on a 7-point Likert scale.
    \textbf{Top}: Human-Likeness. \textbf{Bottom}: Realism.
    Segment width and colour intensity indicate rating frequency (1 = \textit{Not at all}, 7 = \textit{Extremely}).}
    \Description{A horizontal stacked bar chart showing the distribution of peer supporter ratings for human-likeness and realism in Study 1. Each bar represents one evaluation criterion, with segments corresponding to Likert scale values from 1 to 7. Segment width and shading intensity encode the number of ratings at each scale point, making visible the concentration of responses towards the mid-to-upper range of the scale.}
    \label{fig:likert_graphs_study_1}
\end{figure}

\paragraph{\textbf{Perceptions of Naturalness, Realism, and Human-Likeness.}}
Participants reported moderately high perceptions of \textit{realism} and \textit{human-likeness}, with 58\% describing the interaction as natural.
Median ratings were 4 for human-likeness and 5 for realism on a 7-point scale (Figure~\ref{fig:likert_graphs_study_1}).

Qualitative accounts revealed a recurring tension between emotionally credible content and interactional artificiality.
Many participants ($n=7$) found \client{}’s concerns realistic and familiar to their peer support experiences--particularly around academic stress, future uncertainty, and social isolation.
At the same time, several ($n=4$) noted that \client{}’s delivery felt overly fluent, structured, and emotionally regulated, lacking hesitation, digression, or irregularity commonly found in distressed individuals.
As \greenleaf{} (who had 3-4 years of experience supporting others online) observed, \dquote{Real people don't speak in paragraphs...they just blurt things out}.
Others ($n=2$) described \client{} as unusually cooperative or emotionally stable compared to real distressed individuals.

Despite these limitations, some participants ($n=3$) described moments of conversational reciprocity that supported sustained engagement, noting that \client{} occasionally built on prior messages in ways that felt emotionally responsive.
Overall, these reflections suggest that while \client{} lacked key features of human-to-human interaction and was recognisably artificial, its emotional content and relational cues were sufficiently credible to support engagement.

Taken together, these reflections suggest that while \client{}'s delivery lacked features of human-to-human conversation, its emotional content and relational cues were sufficiently credible to sustain engagement with participants.

\subsubsection{Engagement with \suggestions{}}
\begin{figure}[htbp]
  \includegraphics[width=\textwidth]{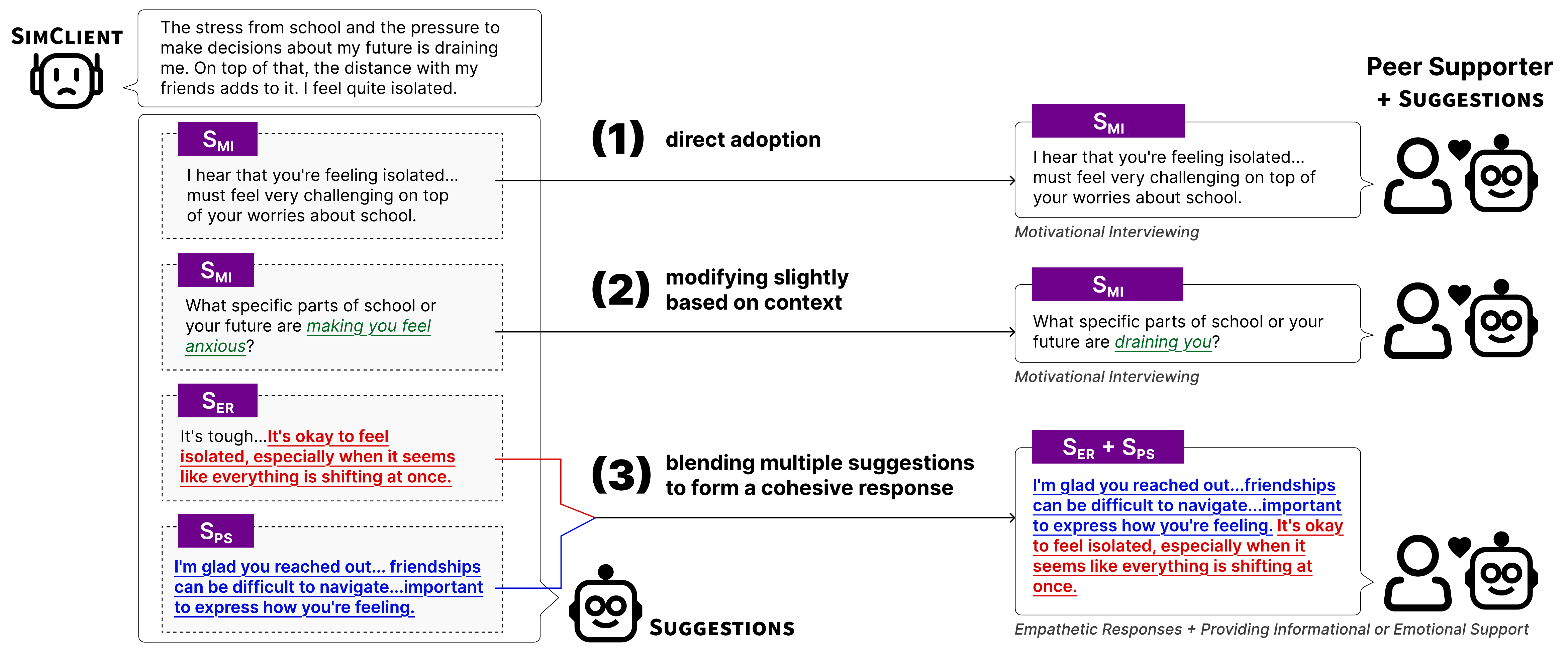}
  \caption{Overview of participants' interactions with \client{}, with the usage of \suggestions{} through three strategies: (1) direct adoption, (2) slight modification, and (3) combining suggestions.}
  \Description{Diagram illustrating participants' interactions with an LLM-simulated client. The figure depicts three strategies for using the LLM-generated suggestions: direct adoption, slight modification, and combining multiple suggestions into a single response.}
  \label{fig:suggestion-usage}
\end{figure}

\begin{figure}[htbp]
    \centering
    \begin{subfigure}[t]{0.48\textwidth}
        \includegraphics[width=\textwidth]{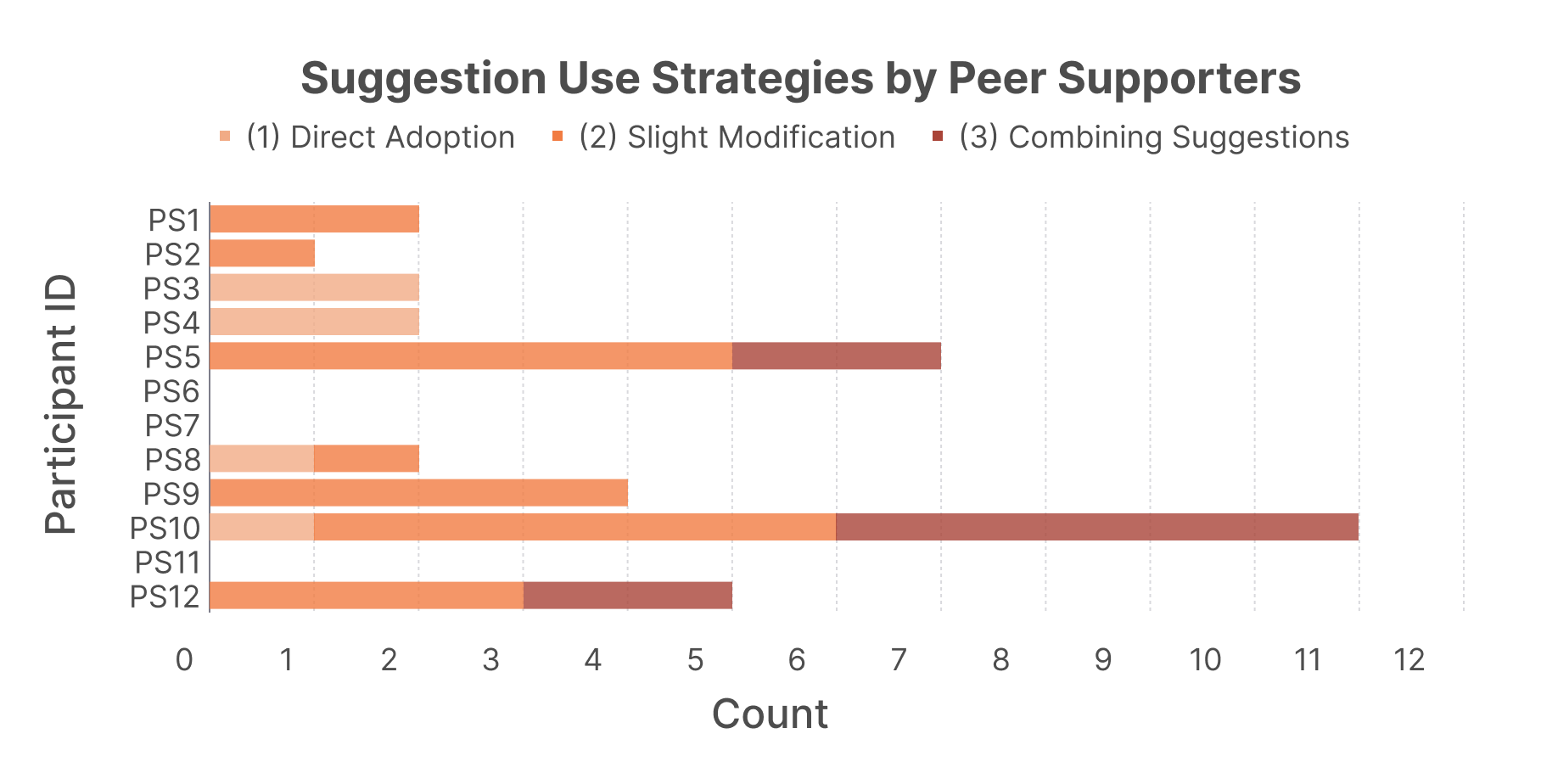}
        \caption{Suggestion use strategies by participant}
        \label{fig:suggestion_use_strategies}
    \end{subfigure}
    \hfill
    \begin{subfigure}[t]{0.48\textwidth}
        \includegraphics[width=\textwidth]{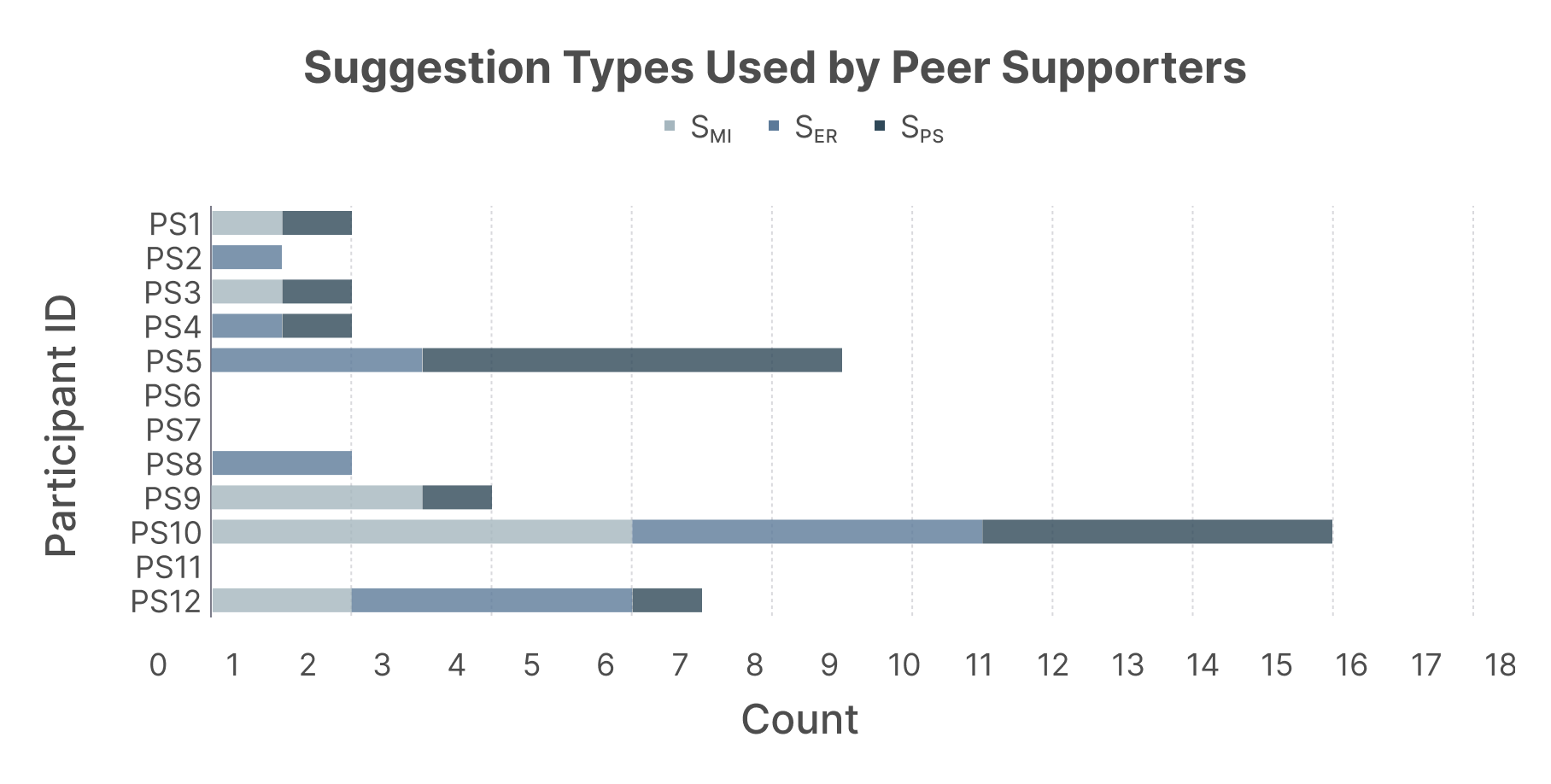}
        \caption{Suggestion types used by participant}
        \label{fig:suggestion_type_used}
    \end{subfigure}
    \caption{Stacked bar charts showing how each participant in Study 1 used \suggestions{}. \textbf{Left}: Usage strategies (direct adoption, slight modification, blending). \textbf{Right}: Suggestion types used, namely motivational interviewing (\smi{}), empathetic responses (\ser{}), and emotional support (\sps{}).}
    \Description{Two horizontal stacked bar charts. The left chart shows for each participant how suggestions were used: directly, slightly modified, or blended. The right chart shows the number of suggestions used for each type: motivational interviewing, empathetic responses, and emotional support. Each bar corresponds to a participant, with distinct colours for each category.}
    \label{fig:suggestion_use_strategies_and_types}
\end{figure}

\paragraph{\textbf{Usage Patterns and Strategies.}} 
9 out of 12 participants used \suggestions{}, with an average adoption rate of 29.1\% ($SD = 23.62\%$).
They used \suggestions{} in three ways (see Figure \ref{fig:suggestion-usage}: (1) direct adoption; (2) modifying slightly (trimming/adding content) based on context; (3) blending multiple suggestions to form a cohesive response. 
Modified use was the most common approach (21 instances), followed by blending (9 instances) and direct adoption (6 instances), as shown in Figure~\ref{fig:suggestion_use_strategies}.
For example, \windsong{} directly adopted suggestions when unsure how to reply, while \brightstar{} and \earnprom{} regularly modified suggestions for clarity, tone, or personal voice. 
\duskytide{} described applying peer support techniques to enhance the phrasing of suggestions, and \earnprom{} often combined elements from different options, preferring the structure provided by \suggestions{} over composing replies from scratch.

Participants gravitated towards \ser{} and \sps{} (16 instances each), using these to establish rapport and emotional containment.
Motivational interviewing prompts (\smi{}) were used less frequently (13 instances) and were typically employed for reflection or reframing rather than directive guidance (see Figure~\ref{fig:suggestion_type_used} (right)).
As \brightstar{} explained, they preferred to \dquote{start with empathy before probing, normalising, affirming, and helping \client{} seek...resolution}.

\paragraph{\textbf{Perceived Usefulness and Limitations.}}  
Seven participants described \suggestions{} as broadly aligned with principles taught during trainings, particularly in modelling empathy, validation, and non-judgemental language.
These participants found all three suggestion categories useful for initiating emotionally sensitive responses, structuring replies, or evaluating drafts they had already composed.

For some, \suggestions{} functioned as confidence scaffolds.
\greenleaf{} highlighted their potential value for novice peer supporters, noting that they helped ensure alignment with fundamental support principles, while \duskytide{} described using them for ideation when unsure how to proceed.
Others, such as \brightstar{} and \earnprom{}, found \suggestions{} helpful for expanding or refining their responses, with \earnprom{} commenting that some suggestions were \dquote{better than what [they] would have typed}.
While \bluemoon{} and \redballoon{} maintained personal styles, such as rapport-building approaches, they drew inspiration from \suggestions{} verbatim.

However, limitations were also noted.
Several participants ($n=4$) felt that \suggestions{} were insufficiently attuned to conversational context, sometimes addressing issues before rapport had been adequately established.
As \duskytide{} observed, \dquote{[\suggestions{}] didn't always match what [\client{}] was actually saying}.
Concerns about authenticity also emerged.
Some participants ($n=3$) described the suggestions as overly generic or \dquote{textbook}, expressing discomfort with language that felt scripted or emotionally loaded too early in the interaction.
\redballoon{} worried that phrases such as \dquote{I'm here for you, you are not alone} could foster co-dependency if used prematurely, while \bluemoon{} preferred gentler approaches such as open-ended questioning or distraction over explicit reassurance.

\paragraph{\textbf{Usability Challenges and Design Suggestions.}}  
Participants reported mixed experiences with the timing and responsiveness of \suggestions{}. 
Several participants ($n=3$) noted that suggestions often appeared too late in the interaction, sometimes only after they had already composed or sent a response.
This latency reduced the practical usefulness of the feature during active message composition and occasionally disrupted conversational flow.

Repetition and limited contextual adaptation were also raised as concerns.
Two participants observed repetition across suggestions and felt that outputs did not sufficiently reflect the evolving chat history.
These participants expressed a desire for suggestions that adapted more clearly to prior turns, emotional shifts, and conversational pacing.

Participants proposed a range of design refinements to address these issues.
Several suggested enabling quicker access to support, such as prompts or commands that could be invoked on demand (\earnprom{}), or offering suggestions earlier in the drafting process.
Others proposed interaction techniques that would allow supporters to tailor suggestions more actively, including drag-and-drop editing (\brightstar{}), thumbs-up or thumbs-down feedback to guide future outputs (\bluemoon{}), and a hybrid approach combining full responses with keyword-level prompts (\windsong{}).
Some participants also suggested extending \suggestions{} beyond response generation.
\greenleaf{} recommended integrating resource links for situations involving heightened risk, while \greenship{} proposed contextual feedback on typed replies informed by professional best practices.
\bluemoon{} suggested allowing \suggestions{} to temporarily take over the conversation to enable short breaks without disrupting support.

\subsubsection{Use of Emotional State Visualisations}
\paragraph{\textbf{Patterns of Use and Interpretation.}}
Participants differed in their engagement with the emotional state visualisations.
Three participants reported active use, primarily consulting the arousal graph to guide conversational pacing and tone.
\zoomplane{} and \greenleaf{} used the graph to confirm when to wrap up the conversation, while \ironpetal{} adjusted their tone upon seeing a \dquote{medium to negative} shift.
Eye-tracking data corroborated this, showing glances at the visualisations before typing responses in key moments of tone shift.

A second group ($n=3$) reported occasional or confirmatory use, glancing at the visualisations but finding the direct chat content more informative, as they felt the graphs were more suited for retrospective analysis.
The remaining participants ($n=6$) reported minimal or no use.
These participants either did not understand how to interpret the visualisation or found it irrelevant to their support strategies, focusing instead on composing responses.
\windsong{} only paid attention to the visualisations mid-session, rating their influence as \dquote{1.5 out of 5 stars}.

Across participants, textual cues were consistently prioritised over graphical indicators.  
Even among active users, the visualisations informed behaviour intermittently rather than continuously, functioning as a secondary reference rather than a primary driver of action.

\paragraph{\textbf{Suggestions for Design Improvements.}}
Participants proposed refinements to improve clarity and usability.
Some ($n=3$) found technical terms such as \dquote{valence} hard to interpret and suggested layperson alternatives like \dquote{engagement level}. 
Others ($n=3$) proposed visualising a small set of dominant emotions over time to better convey emotional trends without increasing visual complexity, or using colour-coded indicators to reduce cognitive load.
Interface-level adjustments were also suggested.
\earnprom{} proposed repositioning the visualisation closer to the \client{} label to improve visibility, while \zoomplane{} preferred replacing the valence-arousal display with more direct representations of basic emotions.
Some participants noted that constantly monitoring the visualisation increased emotional pressure rather than support, describing the feature as both \dquote{stressful} and potentially helpful (\greenleaf{}).
While a few participants found the tool informative or affirming, most ultimately preferred relying on text and interpersonal cues. 
The visualisation was seen as a promising adjunct but needed refinement to better support real-time peer support.

\subsubsection{Individual Differences in Engagement}
To move beyond narrative contrasts, we examined whether variation in \suggestions{} usage could be explained by peer supporters' AI familiarity, personality traits, and years of experience.

\paragraph{\textbf{Familiarity with AI}}
Participants reported moderate familiarity with AI ($mean= 3.75$, $SD = 1.06$ on a 5-point Likert scale) and moderate usage frequency ($mean = 3.83$, $SD = 1.80$) on a 7-point Likert scale).
Notably, 6 participants (50\%) reported using AI tools fewer than 4 times per week.
However, correlations with \suggestions{} usage were weak: familiarity ($r = 0.12$) and usage frequency ($r = 0.11$), suggesting that general AI familiarity may not reliably predict how participants integrate \suggestions{} into their interactions.
While higher-frequency AI users like \brightstar{} and \earnprom{} (6-7 times and 10+ times per week, respectively), adopted \suggestions{} more readily (70\% and 64.71\% respectively), other frequent users like \redballoon{} avoided them entirely, citing stylistic misalignment.
Conversely, \duskytide{} adopted suggestions at a modest rate (15.38\%) despite rating themselves as not familiar (1) and reporting no AI use.
\windsong{}, who rated infrequent usage (2-3 times per week), remarked that \dquote{suggestions will be helpful for someone...just starting out}, especially in moments of uncertainty.

\paragraph{\textbf{Personality Traits.}}
Participants showed moderate levels of Conscientiousness ($mean = 3.17$, $SD = 0.50$ on a 5-point scale), indicating a tendency toward being organised and goal-oriented, though not strongly so. 
Conscientiousness exhibited the strongest association with \suggestions{} usage, though the relationship was weak ($r = -0.25$), possibly reflecting a preference for crafting personal responses. 
Other Big 5 personality traits and the Need for Cognition values showed no meaningful associations (see Appendix~\ref{appendix:phase-2personality-traits}).
While personality traits alone might not predict suggestion adoption, they may interact with support style and perceived system alignment.

\paragraph{\textbf{Physiological Indicators.}}  
Physiological measures showed variability across participants and interaction segments. 
We examined heart rate variability using RMSSD as an index of parasympathetic activity, comparing pre-chat baseline segments with chat interaction segments, as well as variation across shorter and longer response intervals within the conversation.
7 out of 12 participants experienced an increase in RMSSD from the pre-chat to chat phase, while 9 out of 12 showed a decrease in mean heart rate (mean change: $-1.91 bpm$), suggesting that for many, physiological arousal subsided once the conversation began.
Longer interaction segments, operationalised based on the timing between participants' and \client{}'s messages, were associated with higher values on average than shorter segments.
This indicates greater parasympathetic activation during extended periods of engagement, which may reflect sustained attention, reflection, or message composition rather than acute stress responses.
Importantly, physiological patterns did not consistently align with self-reported affect or observed system use.
Participants who described the interaction as calm or emotionally neutral sometimes exhibited elevated RMSSD, while others showed little physiological change despite reporting moments of hesitation or cognitive effort.
Similarly, HRV measures did not systematically differ between participants who frequently used \suggestions{} and those who did not.
Taken together, these findings suggest that HRV variation seemed to reflect differences in engagement dynamics and response formulation rather than subjective distress or system-induced cognitive load.

Eye-tracking data were used to examine visual attention to system components during the interaction, particularly the chat window, \suggestions{}, and the emotional state visualisations.  
Across participants, gaze was typically concentrated on the chat interface during active message composition, with minimal fixation on other system elements while typing.
For participants who reported active use of \suggestions{} or emotional state visualisations, eye-tracking data showed brief but repeated fixations on these elements at specific interactional moments, such as after receiving a client message or while waiting for the next response.  
These fixations typically occurred before drafting a reply or after a message had been sent, rather than during the act of composing text.
One exception was \greenleaf{}, who reported consulting the visualisation to assess whether their responses were helping the client emotionally. 

Participants who reported minimal or no use of these features nonetheless exhibited occasional glances toward them, though these were short and infrequent.
These glances were most commonly observed during idle periods, such as while waiting for \client{}'s response, and did not reliably precede changes in response content or strategy.

Overall, sustained visual attention to \suggestions{} or the emotional state visualisations did not systematically correspond with higher adoption rates or stronger behavioural influence.
This suggests that these features functioned primarily as background references rather than directive inputs, supporting their role as optional scaffolds rather than focal decision-making tools.

\paragraph{\textbf{Years of Experience.}}
Years of peer support experience showed a modest negative association with \suggestions{} usage ($r = -0.24$).
A median split indicated that more experienced peer supporters relied on \suggestions{} less frequently than less experienced participants, suggesting that experience may shape how scaffolding tools are integrated into practice.
\section{Study 2: Experts' Perspectives on Study 1's Interactions}
\subsection{Methodology}

\subsubsection{Participants}
We recruited 6 mental health professionals (4 female), comprising 1 clinical psychologist, 4 licensed counsellors, and 1 medical social worker (Table~\ref{tab:study-2-demographics}).
Participants were aged 25 and above ($mean = 39.5$, $SD = 8.9$), with 4 to 20 years ($mean = 11.6$, $SD = 7.70$) of professional experience, and held formal qualifications in clinical psychology, counselling, or social work, with at least a Bachelor's degree in Psychology, Social Work, or a related field.
Outside of their professional careers, two participants (\velvetecho{}, \mistybloom{}) also volunteered as a peer supporter on online platforms.
The medical social worker in this study (\stormvale{}) held advanced certifications and was actively engaged in delivering mental health interventions alongside broader medical social services.
While these participants bring formal psychosocial training, their evaluations are treated as perspectives grounded in professional therapeutic norms rather than a definitive benchmark of correctness.
Given the lack of standardised definitions in peer support, this study aims to identify points of convergence and divergence between professional and peer support practices, rather than a fixed standard.

\begin{table*}[htbp!]
\centering
\small
\caption{Study 2 Participant Characteristics}
\label{tab:study-2-demographics}
\renewcommand{\arraystretch}{1.2}
\begin{tabular}{p{0.25cm} p{0.7cm} p{0.8cm} p{1.2cm} p{1.25cm} p{1.6cm} p{2cm} p{1.8cm} p{1.45cm}}
\toprule
\textbf{ID} & \textbf{Age} & \textbf{Gender} & \textbf{Education} & \textbf{Ethnicity} & \textbf{Employment} & \textbf{Occupation} & \textbf{Peer Support \newline Volunteering \newline (Outside Work)} & \textbf{Years of \newline Professional Experience} \\
\midrule
\stormvale{}   & 25-34 & Female & Bachelor's & Chinese & Contract or \newline temporary & Medical \newline Social Worker & No  & 6  \\
\velvetecho{}  & 45-54 & Male   & Master's   & Chinese & Part-time             & Counsellor            & Yes & 5  \\
\azureflame{}  & 45-54 & Female & Bachelor's & Chinese & Full-time             & Counsellor and \newline Social Worker & No & 20 \\
\twilightrush{}& 35-44 & Male   & Doctoral   & Indian  & Full-time             & Counsellor            & No  & 7  \\
\mistybloom{}  & 35-44 & Female & Master's   & Chinese & Full-time             & Clinical \newline Psychologist & Yes & 20 \\
\solarwisp{}   & 25-34 & Female & Master's   & Malay   & Full-time             & Counsellor            & No  & 4  \\
\bottomrule
\end{tabular}
\end{table*}

\subsubsection{Study Procedure}
Each session comprised five segments: Pre-Interview, Video Review 1, Post-Video Interview 1, Video Review 2, and Post-Video Interview 2, all conducted in person and screen- and audio-recorded.

\paragraph{\textbf{Pre-Interview Segment.}}
Participants completed a demographics questionnaire and a semi-structured interview on their professional background, experiences providing psychosocial support, and perspectives on digital mental health interventions. 
This provided contextual grounding for subsequent evaluations.

\paragraph{\textbf{Video Review and Think-Aloud Segments.}}
Participants then reviewed two anonymised recordings from Study 1 (Chat Segment), selected to vary in duration and use of \suggestions{} and emotional state visualisations (Appendix~\ref{appendix:phase-3-expert-video-mapping}), and provided concurrent think-aloud reflections on conversational appropriateness, realism, and the peer supporter's decision-making processes from a professional perspective.

\paragraph{\textbf{Post-Video Questionnaire and Interview Segments.}}
Following each video, participants completed a short questionnaire evaluating \client{}, \suggestions{}, and the emotional state visualisations. 
This was followed by a semi-structured interview on the peer supporters' responses, strengths and limitations, and the LLMs' role in shaping the interactions.
After answering questions about the second video, participants offered overarching reflections on the potential and pitfalls of LLM-supported peer support.

\subsubsection{Data Analysis}
Similar to Study 1, a single researcher iteratively coded the interview and think-aloud transcripts, refining codes and themes through repeated close readings and memo writing.
\subsection{Results}
\subsubsection{Evaluation of Peer Supporter Responses}
\paragraph{\textbf{Message Length, Conversational Pacing, and Interactional Quality.}}
To contextualise expert evaluations, we examined basic conversational characteristics, including session duration, number of turns, and message length distributions (Table~\ref{tab:study-1-session-stats}; Figure~\ref{fig:phase2-average-message-length}).
However, expert interpretations of pacing and interactional quality were not directly reducible to these surface-level measures.

Experts consistently evaluated pacing as an interactional quality, shaped by how conversational space was held through turn sequencing, responsiveness to \client{}'s cues, and the timing of reassurance or advice.
Shorter interactions were not perceived as rushed, nor were longer sessions necessarily experienced as more reflective.
For instance, \twilightrush{} described \zoomplane{}'s interaction, one of the shortest sessions (approximately 20 minutes, 34 turns), as moving too quickly.
Similarly, \windsong{}'s session (approximately 23 minutes, 24 turns) was critiqued by \velvetecho{} for failing to hold space and prematurely orienting toward closure, despite comparable message lengths and turn counts to other sessions.
Conversely, substantially longer sessions with higher turn counts (e.g., \bluemoon{}, \redballoon{}) were also critiqued, but for different pacing-related issues, including cognitive overload and conversational drift rather than speed.
Together, these cases indicate that expert perceptions of pacing were driven less by elapsed time or verbosity, and more by interactional control and emotional attunement.

Premature advice-giving was observed across both brief and long interactions, indicating that advice-giving reflected peer supporters' conversational strategies rather than cumulative interaction length.
In longer sessions, this sometimes manifested as increasingly directive guidance, whereas in shorter sessions, advice was introduced before sufficient exploration of \client{}'s concerns.
Experts further noted that advice-giving was frequently compounded by assumption-making, where peer supporters inferred emotions, motivations, or problem structures not explicitly articulated by \client{}, sometimes introducing language misaligned with \client{}'s framing.

\paragraph{\textbf{Observed Strengths and Gaps.}}  
Experts highlighted strengths across peer supporter responses, including the use of PFA- and counselling-aligned techniques such as paraphrasing, affirmation, and normalisation.
When grounded in \client{}'s language and used to invite elaboration, these strategies were generally experienced as supportive and emotionally attuned.
For example, \stormvale{} noted that \duskytide{} responded to \client{}'s comment about enjoying social aspects of gaming with a mirrored prompt inviting elaboration, while affirmation and normalisation were observed when in responses such as \dquote{Those sound like concrete suggestions} (\stormvale{} on \duskytide{}) and {Yes, it is indeed normal} (\azureflame{} on \greenship{}, when \client{} questioned the legitimacy of their feelings).
\azureflame{} also commended \brightstar{} for encouraging \client{}'s openness with \dquote{I appreciate your bravery in opening up to me}.

Experts also highlighted the effective use of empathetic statements grounded in personal insight without shifting focus from \client{}.
For example, \azureflame{} commended \brightstar{} for using personal strategies to create resonance without overshadowing \client{}, creating a sense of shared experience. 
\velvetecho{} commended \windsong{} for using \dquote{I totally get how you're feeling}, sharing that the \dquote{use of \enquote*{I} statements} could effectively convey empathy while showing readiness to listen.
More informal elements, such as greetings, emojis, and natural typographic variation, were also seen to contribute to a more human and relatable tone (\solarwisp{} on \redballoon{}).

Despite these strengths, experts observed several interactional gaps.
Scripted or overly formal validation was frequently described as mechanical, reducing emotional authenticity despite conveying basic empathy.
For example, \stormvale{} commented that while \amberlight{}'s tone was less robotic overall, it occasionally strayed from best practices, and that \duskytide{}'s repeated use of phrases such as \dquote{Thanks for sharing} felt formulaic rather than validating.

Experts also raised concerns about linguistic accessibility and cognitive load.
Responses perceived as verbose, abstract, or linguistically complex were described as cognitively demanding, particularly in emotionally salient moments (\stormvale{}, \twilightrush{}, \mistybloom{}).
These critiques were not solely tied to message length, but to sentence structure, abstraction, and the introduction of concepts not grounded in \client{}'s language, risking reduced emotional attunement even when intentions were supportive.

\paragraph{\textbf{Tensions and Misalignments with Professional Standards}}
Experts consistently noted that while many peer supporter responses were well-intentioned, they occasionally diverged from professional counselling expectations, which they believed could be due to differences in training, role clarity, and situational judgement.
Premature advice-giving was common, particularly before fully understanding \client{}'s concerns.
This tendency appeared across multiple sessions and took several forms.
In some cases, peer supporters like \brightstar{} introduced actionable suggestions too early, limiting opportunities for co-regulation and reflection.
For example, \twilightrush{} critiqued \bluemoon{} for shifting from distraction to advice-heavy replies that conflicted with peer support principles, and also described the \zoomplane{}-\client{} interaction as resembling coaching rather than PFA-oriented support, due to the limited exploratory depth despite the provision of resources.
Similarly, \mistybloom{} observed that \greenleaf{} sometimes offered quasi-therapeutic advice, even if this was occasionally softened by checking whether such suggestions would be helpful.

Misalignment also arose in how peer supporters interpreted and acted on \client{}'s cues.
Experts noted instances where emotions, motivations, or progress were inferred without sufficient grounding in \client{}'s expressed experiences.
For example, responses that introduced personal framing or assumed progress were seen as misaligned with \client{}'s emotional state. 
\azureflame{} observed that \greenship{}'s remark, \dquote{I am just like you too. I would find people to talk to express my worries}, felt disconnected from \client{}'s low energy and missed the opportunity for deeper reflection.
Likewise, \velvetecho{} suggested that \windsong{} may have mistaken \client{}'s agreement for genuine progress, overlooking possible people-pleasing tendencies.

A related concern involved responses that redirected or diluted emotional processing rather than holding space for it.
Some interactions became increasingly interrogative (e.g., the \butterflywave{}-\client{} interaction) or shifted towards distraction (e.g., the \bluemoon{}-\client{} interaction), resembling structured questioning or topic diversion rather than sustained emotional engagement.
Specifically, \twilightrush{} noted that \bluemoon{} tried to distract \client{} by asking \client{} to focus on leaves turning yellow.
While such strategies may temporarily reduce distress, experts cautioned that they risk suppressing rather than supporting emotional processing.

These concerns became especially salient in moments of possible risk or heightened vulnerability.
In such cases, moving too quickly towards solutions without adequately probing underlying distress was seen as a critical limitation, particularly where potential risk cues were present (e.g., \client{}'s comment to \amberlight{} about wanting to \dquote{escape from all these}).

\paragraph{\textbf{Cognitive Load and Interactional Sustainability.}}
Experts frequently identified cognitive load as a key challenge, with 10 interactions evaluated as overly demanding for \client{}.
These concerns were not solely driven by message length, but by how complexity, questioning, and conversational pacing were managed.

Responses that combined multiple prompts, abstract phrasing, or directive guidance were often experienced as overwhelming.
For example, \solarwisp{} noted that \redballoon{}'s use of double questions or prompts (e.g.,\dquote{How much do you think they value this friendship?}) and prolonged small talk lasting more than 10 minutes led to a \dquote{pen-pal conversation}, delaying emotional disclosure and increasing cognitive burden. 
Similarly, \mistybloom{} noted that \greenleaf{}'s responses, while well-intentioned, risked being overwhelming due to their length and complexity.

Experts also highlighted how cognitive load was shaped by interactional timing and readiness.
Directive suggestions introduced by \redballoon{} without first establishing \client{}'s readiness were seen as particularly demanding, as they centred on peer supporter assumptions rather than \client{}'s needs.
Similarly, \twilightrush{} observed that \bluemoon{}'s long and suggestive messages imposed excessive cognitive load, remarking that there were \dquote{too many words to take in} given the emotional context.
In one instance, \bluemoon{}'s tone was described as so scripted and disconnected that \twilightrush{} questioned whether the peer supporter was human, almost akin to an AI-generated peer supporter planted to test their judgement.

\paragraph{\textbf{Expert Critiques of Supporter Judgement.}}
Experts frequently referenced the presence of \suggestions{} when evaluating peer supporter responses to assess how supporters exercised judgement in adopting, adapting, or disregarding them.
Rather than evaluating correctness, experts focused on whether supporter choices reflected responsiveness to \client{}’s cues, conversational timing, and alignment with peer support principles.

\begin{figure}
    \centering
    \includegraphics[width=\linewidth]{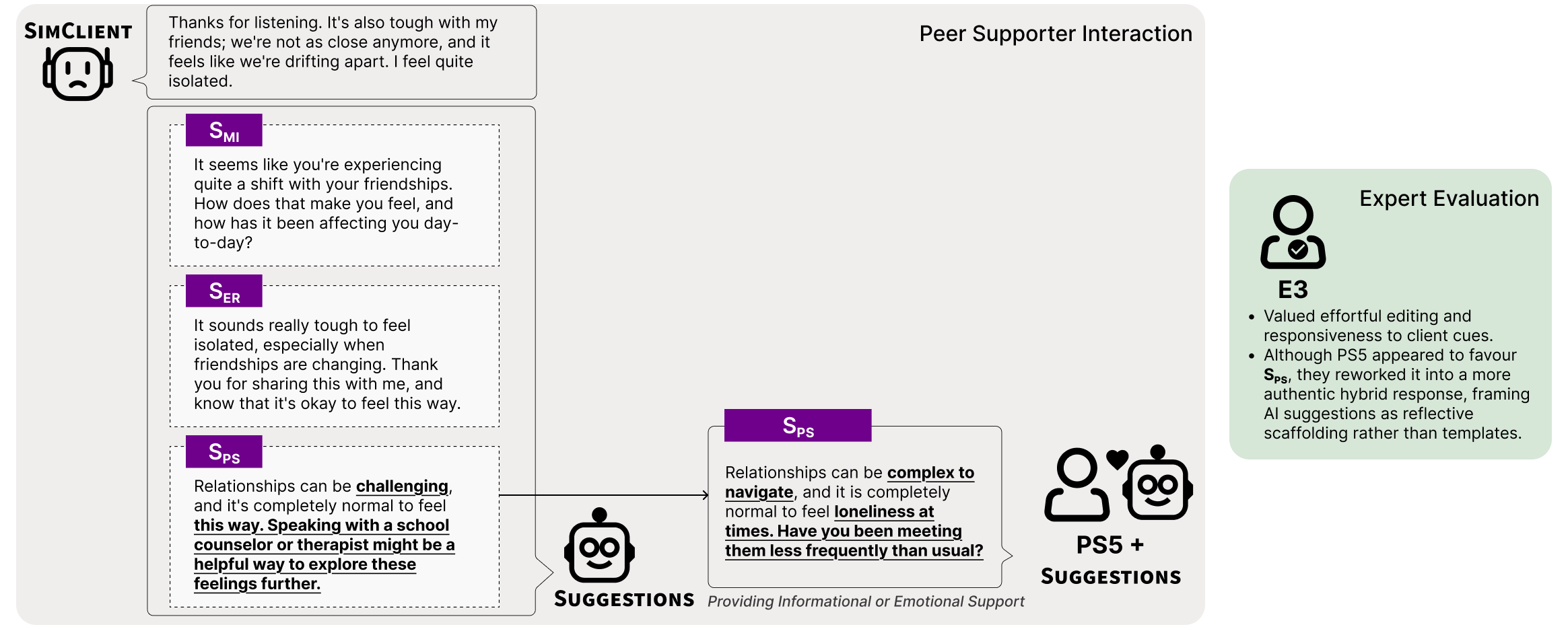}
    \caption{Annotated example of a peer supporter interaction with \client{}, illustrating how \suggestions{} was selectively adapted into an empathetic, reflective response and evaluated by domain experts as scaffolding rather than prescriptive templates.}
    \Description{Annotated diagram showing a text-based peer support conversation with a simulated client who expresses isolation and drifting friendships.
    On the left, the client message is followed by three AI-generated suggestion types labelled S_MI, S_ER, and S_PS, representing reflective questioning, emotional validation, and problem solving or support.
    The peer supporter integrates and edits these suggestions into a single hybrid response that reframes loneliness as normal and asks a gentle follow-up question.
    On the right, an expert evaluation panel highlights that the peer supporter prioritised effortful editing and responsiveness, transforming the AI output into authentic reflective scaffolding rather than using it verbatim.}
    \label{fig:study-2-annotated-example}
\end{figure}

Across sessions, experts consistently valued selective and effortful adaptation over verbatim adoption (see Figure~\ref{fig:study-2-annotated-example}).
When peer supporters reworked available suggestions to better fit conversational context, tone, or sequencing, this was interpreted as evidence of skilled judgement rather than deviation.
\suggestions{} functioned as a comparative reference point rather than a prescriptive guide, and expert critiques centred on how supporters navigated trade-offs between structure and flexibility, using suggestions as scaffolding while maintaining ownership over the interaction.
Annotated examples illustrating these judgement patterns are provided in Appendix~\ref{appendix:phase-3-annotated-examples}.

Experts also highlighted inherent role tensions in peer support.
Given the non-clinical setting, peer supporters were not expected to perform therapeutic tasks, such as formal risk assessment or structured problem-solving.
Accordingly, behaviours such as encouragement statements, informal reassurance, and limited self-disclosure were viewed as acceptable within peer support contexts, even when misaligned with professional counselling norms (\stormvale{}, \azureflame{}, \mistybloom{}).
These were interpreted as forms of boundary negotiation, in which peer supporters balanced relational warmth with the risk of becoming overly directive or invalidating.
However, they also discouraged the use of vague encouragement (e.g., \dquote{you should give yourself credit}), overgeneralised phrases (e.g., \dquote{I understand}), or abrupt conversational endings without clear summarisation, which they warned could undermine trust and leave emotional needs insufficiently contained.

\paragraph{\textbf{Recommendations for Improvement}}
Experts offered recommendations to enhance the quality, safety, and emotional efficacy of peer support interactions. 
First, experts emphasised the importance of warmer, more affirming openings to reduce hesitation and encourage disclosure.
For instance, \velvetecho{} suggested reframing \earnprom{}'s opening from \dquote{what brings you to us today?} to \dquote{hello, welcome...your presence here matters} in order to lower barriers to disclosure.

Second, experts highlighted the need to calibrate self-disclosure and empathy.
Brief \squote{I} or \squote{we} were encouraged to convey empathy without centring the peer supporter’s experience, as opposed to full personal disclosures (\velvetecho{}), alongside stronger empathetic grounding before follow-up questions (\mistybloom{}).
Such a statement could look like \dquote{I'm really glad you reached out; friendships can be tough}, which could be used to show empathy without centring a peer supporter's personal life.

Third, pacing and conversational focus were identified as critical.
Experts recommended slowing advice-giving, prioritising reflective listening, and de-emphasising problem resolution.
Open-ended questions, paraphrasing, and gentle invitations for \client{} to elaborate were consistently encouraged.
For instance, moving too quickly to problem-solving was seen to limit opportunities for deeper engagement with \client{}'s concerns (\azureflame{}).

Relatedly, experts emphasised the importance of holding space and allowing \client{} to process emotions.
Long, complex, or prematurely directive responses were seen as overwhelming, particularly when \client{} was not yet ready to engage with suggestions (\twilightrush{}, \mistybloom{}).
Even when self-disclosure was appropriate, it needed to remain clearly anchored to \client{}'s experience to maintain relevance (\solarwisp{}).
This would help to ensure that the peer supporter's responses do not impose additional cognitive or emotional overload on \client{}.

Finally, experts highlighted the role of intentional and emotionally contained closure.
They recommended summarising key themes, acknowledging emotional effort, and signalling availability for continued support.
For example, vague closings (e.g., \dquote{Peace be with you} by \bluemoon{}) were critiqued by \twilightrush{} as insufficiently grounding, while more structured endings (e.g., by clarifying progress, providing closure, signalling availability for follow-up) were seen by \twilightrush{} and \solarwisp{} to better support continuity and trust.
To better align with peer support principles, rather than offering unsolicited tips, \solarwisp{} recommended reframing suggestions into probing, client-centred prompts, such as \dquote{What would this look like for you, if you were to explore your strengths in this area?}.

\subsubsection{Evaluation of LLM-Supported Components}
\begin{figure}[htbp]
    \centering
    \includegraphics[width=0.9\textwidth]{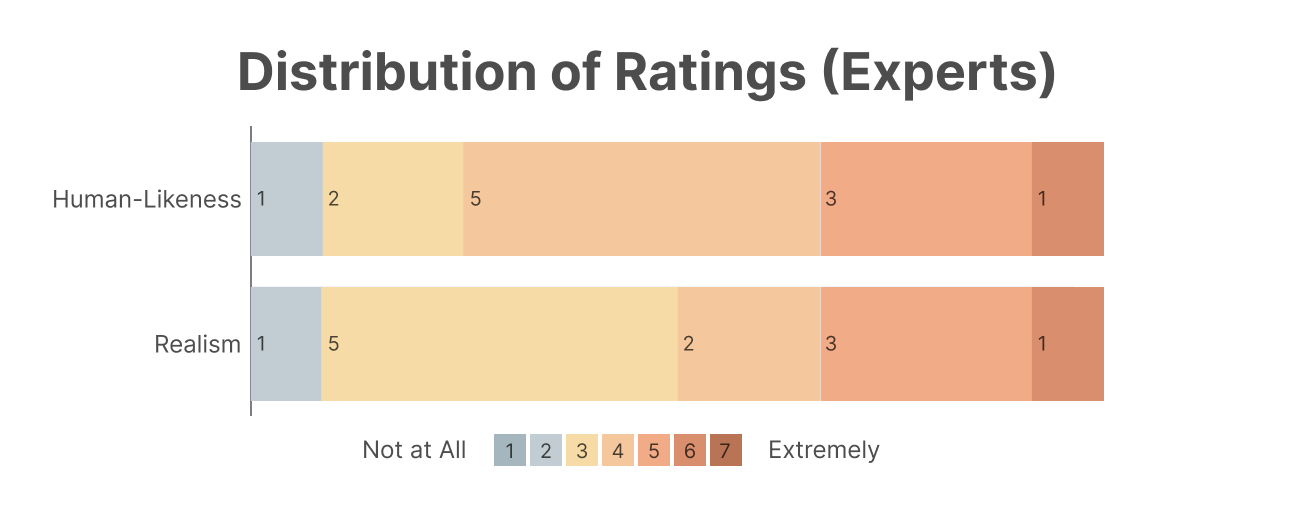}
    \caption{Distribution of expert ratings of \client{} in Study~2 across a 7-point Likert scale.
    Segment width and colour intensity indicate the frequency of ratings at each scale point.
    \textbf{Top}: Human-Likeness. \textbf{Bottom}: Realism.
    Ratings range from 1 (\textit{Not at all}) to 7 (\textit{Extremely}).}
    \Description{A horizontal stacked bar chart showing the distribution of expert ratings for human-likeness and realism in Study 2. Each bar represents one evaluation criterion, with segments corresponding to Likert scale values from 1 to 7. Segment width and shading intensity encode the number of expert ratings at each scale point, highlighting greater concentration and agreement for realism compared to human-likeness.}
    \label{fig:likert_graphs_study_2}
\end{figure}

\paragraph{\textbf{Evaluation of \client{}.}}
Expert evaluations revealed a mixed appraisal of \client{}.
Of 12 sessions reviewed, 66.7\% were rated as \textit{natural}.
Median ratings were 4 for \textit{human-likeness} ($IQR = 1.25$) and 3.5 for \textit{realism} ($IQR = 2$), as shown in Figure~\ref{fig:likert_graphs_study_2}.

Alignment between expert and peer supporter ratings was weak (Cohen’s $\kappa = -0.24$ for \textit{naturalness}; $r = 0.20$ for \textit{human-likeness}; $r = 0.18$ for \textit{realism}), indicating disagreement despite the use of identical rating scales.
As shown in Table~\ref{tab:peer_expert_ratings}, 11 of the 12 sessions exhibited at least one point of disagreement.
Peer supporters consistently rated \client{} more favourably than experts, particularly on subjective dimensions such as \textit{human-likeness} and \textit{realism}, with discrepancies ranging from one to four scale points.
\begin{table}[htbp]
    \centering
    \small
    \caption{Peer Supporter (PS) versus Expert (E) ratings across three evaluation criteria from Studies 1 and 2, respectively. Red text indicates disagreement between the PS and E rating for that criterion.}
    \Description{Table comparing peer supporter and expert evaluations of simulated peer support conversations. Each row lists a peer supporter and expert pair, followed by their ratings on naturalness, human-likeness, and realism. Disagreements between peer supporter and expert ratings are shown in red text. \butterflywave{} and \redballoon{} were not included in this table as their sessions had not been evaluated by experts at the time of writing.}
    \begin{tabular}{p{3cm} p{1.8cm} p{2.3cm} p{3cm} p{2cm}}
        \toprule
        \textbf{Peer Supporter (PS)} & \textbf{Expert (E)} & \textbf{Naturalness (PS/E)} & \textbf{Human-Likeness (PS/E)} & \textbf{Realism (PS/E)} \\
        \midrule
        \duskytide{} & \stormvale{} & Yes/Yes & \disagree{4/5} & \disagree{4/5} \\
        \zoomplane{} & \twilightrush{} & \disagree{Yes/No} & \disagree{6/3} & \disagree{6/3} \\
        \windsong{} & \velvetecho{} & Yes/Yes & 4/4 & 5/5 \\
        \amberlight{} & \stormvale{} & Yes/Yes & \disagree{6/5} & \disagree{6/5} \\
        \brightstar{} & \azureflame{} & \disagree{No/Yes} & \disagree{4/5} & \disagree{5/4} \\
        \butterflywave{} & \solarwisp{} & \disagree{Yes/No} & \disagree{4/2} & \disagree{4/2}\\        
        \bluemoon{} & \twilightrush{} & No/No & \disagree{4/3} & \disagree{4/3} \\
        \greenship{} & \azureflame{} & \disagree{No/Yes} & \disagree{2/4} & 3/3 \\
        \ironpetal{} & \mistybloom{} & \disagree{Yes/No} & \disagree{7/4} & \disagree{7/3} \\
        \earnprom{} & \velvetecho{} & Yes/Yes & 6/6 & \disagree{5/6} \\
        \redballoon{} & \solarwisp{} & \disagree{No/Yes} & \disagree{5/4} & \disagree{4/3}\\
        \greenleaf{} & \mistybloom{} & \disagree{No/Yes} & 4/4 & \disagree{5/4} \\
        \bottomrule
    \end{tabular}
    \label{tab:peer_expert_ratings}
\end{table}

Experts appreciated that \client{} raised familiar issues like school stress, friendship difficulties, feelings of isolation, and the desire for emotional validation.
Across sessions, \client{} was described as courteous and appreciative, frequently expressing gratitude (e.g., \dquote{thanks for listening}, \dquote{your support means a lot}) (\stormvale{}, \azureflame{}, \mistybloom{}).
\stormvale{} further observed that \client{}'s orientation towards actionable next steps mirrored tendencies seen among real-world clients seeking concrete guidance.

However, some experts noted that \client{} lacked the hesitation, resistance and rumination typical of real distress, and were instead overly detailed (\azureflame{}, \twilightrush{}, \mistybloom{}).
\client{} often followed the peer supporter's conversational lead (\velvetecho{}, \twilightrush{}) and introduced multiple concerns too rapidly without sufficient emotional anchoring.
As a result, \velvetecho{} noted an over-reliance on cognitive descriptors (e.g., \dquote{anxiety}) rather than affective states (e.g., \dquote{fear}), blurring whether \client{} was experiencing emotional discomfort or cognitive stress.
\twilightrush{} observed that \client{}’s message complexity often mirrored that of \bluemoon{}, producing sophisticated replies to elaborate inputs and simplified responses to brief ones.
Most experts ($n=5$) further remarked that \client{}'s replies were overly polished and grammatically consistent across 9 of the 12 sessions, reducing perceived human-likeness.
\twilightrush{} characterised \client{} as \dquote{too perfect} and overly accommodating, while \mistybloom{} described it as \dquote{too helpful} and implausibly easy to calm.
They noted that real clients would be more likely to dwell on feelings, hesitate, or resist suggestions rather than becoming \dquote{positive and happy} within 11 minutes. 
\solarwisp{} similarly noted that in the interaction with \butterflywave{}, \client{} appeared unrealistically forward, explicitly asking \dquote{Can you help?} and accepting the first suggestion with little resistance instead of dwelling on emotions, hesitating, or resisting support.
\solarwisp{} expected real distressed clients to avoid becoming \dquote{positive and happy} within 11 minutes.
In the \redballoon{} session, \client{} was described as unusually compliant and verbose, at times reversing roles by asking the peer supporter how they were doing or whether they had anything to share.
\velvetecho{} and \azureflame{} interpreted this role inversion as reinforcing an overly agreeable and people-pleasing presentation that obscured deeper emotional distress.

\paragraph{\textbf{Evaluation of \suggestions{}.}}  
Overall, experts rated \suggestions{}'s alignment with PFA or peer support principles as slightly above neutral ($mean = 4.1$ on a 7-point Likert scale, where 1 = Not aligned at all and 7 = Completely aligned).
In forming these evaluations, experts also considered eye-tracking data from Study 1, which showed that some participants visually engaged with \suggestions{} even when not adopted.
This was interpreted as indicative of selective cognitive processing, raising concerns about timing and perceived relevance.

The categorisation of suggestions into distinct types was consistently valued as a form of scaffolding, particularly for less experienced peer supporters.
Experts described this structure as reflective of evidence-based practice (\stormvale{}, \mistybloom{}, \solarwisp{}) and supportive of deliberation (\stormvale{}), allowing supporters to pause and consider alternative phrasings.
\solarwisp{} added that \smi{} and \ser{} could be useful at any point, though \sps{} typically felt more appropriate later in the conversation rather than as an opening move.
In some cases, \suggestions{} was seen as offering more emotionally attuned or relevant options than the peer supporter's replies, such as when \twilightrush{} observed that \bluemoon{} ignored more suitable suggestions.
This was also observed in both the \butterflywave{} and \redballoon{} sessions, where \solarwisp{} remarked that \suggestions{} were generally sound and, at times, more consistent with counselling practice than the peer supporters' responses.
Nevertheless, experts also emphasised that effective use depended on selective adaptation rather than direct adoption, with flexible integration seen as a marker of strong judgement (e.g., \mistybloom{} on \ironpetal{}).

However, experts also identified limitations.
Repetition was a common concern, particularly in the \sps{} category, where solution-oriented prompts would provide informational support and rush into solution-giving without first acknowledging or holding space for \client{}'s emotions (\velvetecho{}, \mistybloom{}).
Some suggestions appeared to be \dquote{half-correct}, offering general empathy or motivational statements but lacking clear linkage to \client{}'s immediate concerns or language (\twilightrush{}).
\velvetecho{} further noted the cognitive load placed on peer supporters, who had to process multiple suggestions, assess contextual fit, and integrate these into their own words, potentially creating confusion in time-sensitive contexts.

Temporal misalignment further shaped these limitations.
\solarwisp{} observed that \suggestions{} could become repetitive or overly static, remaining unchanged across multiple turns in the \butterflywave{} session and drifting into generalities in the \redballoon{} session.
Although updates occurred, they were not always well-tuned to \client{}'s disclosures.
In one case, an empathetic statement such as \dquote{it is part of growing} risked invalidating \client{}'s concerns by normalising them too quickly.

Quantitatively, suggestion use showed a complex relationship with expert evaluations.
Sessions rated as natural by experts exhibited higher average \suggestions{} usage (29.58\%) than those rated as not natural (8.14\%).
Usage was moderately positively correlated with expert-rated naturalness ($r = 0.435$), for both direct ($r = 0.420$) and indirect or merged use ($r = 0.308$).
In contrast, usage was moderately negatively correlated with expert-rated PFA alignment ($r = -0.331$), particularly for indirect or merged use ($r = -0.385$).
Notably, minimal use did not result in more natural interactions.

\paragraph{\textbf{Usefulness of Emotional State Visualisations.}}  
Experts generally interpreted emotional state visualisations (viewed alongside the chat logs) as peripheral rather than central to peer supporters' decision-making.
Although eye-tracking data showed occasional gaze shifts toward the visualisation, these moments were brief and typically occurred during conversational pauses rather than active message composition, suggesting limited sustained engagement.

Consistent with this interpretation, most experts did not comment on the emotional state visualisation unless gaze deviations were particularly noticeable, and even then, framed these as momentary shifts away from the chat interface rather than meaningful interaction.
Quantitatively, experts expressed cautious optimism about the emotional state visualisations: although they were rated helpful in only 40\% of chat logs, all experts saw potential for improving emotional attunement and pacing.

Where engagement did occur, experts highlighted the usefulness of visual trends rather than discrete emotional states.
Experts noted that changes in arousal, for instance, could signal shifts in conversational impact even when valence remained negative.
For example, \stormvale{} noted that reductions in arousal could signal progress even when sadness persisted, while \azureflame{} suggested that improvements in top-bar indicators may have influenced conversational pacing.
However, experts also questioned the realism and reliability of these signals.
\solarwisp{} noted that while arousal trends were visible and sometimes interesting, transitions appeared overly neat compared to real fluctuating emotional dynamics, and \twilightrush{} cautioned that improvements in displayed emotion could misrepresent conversational quality without additional context.
This was observed when \bluemoon{} provided suboptimal responses to \client{}, which were expected to result in annoyance instead of an overly improved valence/arousal shift.

These concerns were compounded by issues of interpretation and usability.
\mistybloom{} and \solarwisp{} warned that untrained peer supporters might over-rely on or misinterpret emotion labels, potentially leading to inaccurate or invalidating responses.
At the same time, the demands of real-time interaction raised questions from \azureflame{} about whether peer supporters could meaningfully engage with the visualisation alongside ongoing conversation.

As such, recommendations focused on improving interpretability and integration.
Several refinements were proposed, including simpler terminology (e.g., \dquote{stress level} instead of \dquote{valence}), clearer emotional labels (e.g., \dquote{anger, fear}), more intuitive layouts, and risk indicators (\stormvale{}, \velvetecho{}, \mistybloom{}, \solarwisp{}).
For real clients, \velvetecho{} and \twilightrush{} proposed combining multiple data points, such as physiological data, keystroke patterns or gaze data, but emphasised that these indicators would need to be baseline-calibrated and continually refined through system feedback.

Taken together, these findings suggest that experts viewed the current implementation of the emotional state visualisations as ambient cues rather than actionable tools, with their value dependent on improved interpretability, validity, and integration into conversational flow.

\subsubsection{Implications for System Design and Training}
\paragraph{\textbf{System Design Considerations.}}
Experts emphasised that system-level design choices in such LLM-supported systems strongly shaped peer supporters' interactional behaviour and cognitive load.
While tools such as \client{}, \suggestions{}, and emotional state visualisations were seen as promising,  their current configurations required refinement to better align with peer support principles.

For \client{}, experts highlighted the need for more \textit{human-like} and \textit{realistic} behavioural variation.
Current interactions were perceived as overly articulate, emotionally contained, and highly cooperative, limiting opportunities for peer supporters to practise managing hesitation, ambiguity, and resistance.
This reduced opportunities to hold space during hesitation or emotional uncertainty, and produced unrealistically smooth conversational flow (\stormvale{}, \twilightrush{}, \mistybloom{}).
To address this, experts proposed introducing greater behavioural variability, including unprompted emotional disclosures, topic shifts, and selective resistance.
For instance, \stormvale{} suggested incorporating spontaneous disclosures, while \solarwisp{} emphasised slowing reply pacing to introduce conversational friction.
Experts also highlighted the importance of grounding \client{} in a more coherent and human-like profile, such as through personality traits, formative experiences, and less polished linguistic patterns (\azureflame{}, \twilightrush{}, \mistybloom{}).

For \suggestions{}, concerns centred on repetition, timing, and cognitive load.
Experts noted that displaying multiple suggestion categories simultaneously required peer supporters to assess relevance, timing, and phrasing under emotional pressure, increasing cognitive demand (\velvetecho{}, \azureflame{}, \twilightrush{}).
Rather than increasing the number of options, experts recommended more context-sensitive delivery.
This included sequencing suggestions according to conversational stage, foregrounding empathetic responses early, and introducing solution-oriented prompts more cautiously.
They also proposed allowing users to control when suggestions are generated, offering multiple phrasing variants within a single category to support more flexible integration, and offering rephrasings within a single category to support more flexible integration (\twilightrush{}, \mistybloom{}, \solarwisp{}).

At the interface level, experts highlighted opportunities to better support peer supporters' orientation, role clarity, and affective awareness.
For example, \stormvale{} suggested using distinct colours or labels to help peer supporters distinguish between multiple clients if needed, while others more broadly emphasised reinforcing the non-professional role of peer supporters, which involves offering emotional support rather than formal therapeutic interventions.
Mechanisms such as checking alignment with \client{}’s emotional state were seen as ways to maintain attunement without overstepping.

Finally, to reduce peer supporter burnout and support their affective attunement, \azureflame{} proposed extending unobtrusive emotion-tracking tools, such as typing patterns and message content, to peer supporters.
\twilightrush{} added that integrating elements, such as tracking valence over time, could be valuable if validated carefully and refined iteratively.
This might allow AI systems to support not only client wellbeing but also the evolving emotional and cognitive states of peer supporters.

\paragraph{\textbf{Implications for Peer Supporter Trainings.}}
Experts viewed the system as a potentially useful training tool, but emphasised that its current design limits its effectiveness for preparing peer supporters for real-world interactions.
They suggested improvements to effectively simulate complex emotional dynamics, client resistance, and human-like distress.

For \client{}, the lack of behavioural complexity was seen as a key limitation.
Experts noted that overly cooperative and emotionally consistent responses reduced opportunities to practise attunement, boundary-setting, and managing distress escalation.
They argued that the system could serve as a first-stage training tool, allowing peer supporters to practise skills such as paraphrasing and summarisation, but would be insufficient preparation for real-world clients who may be dysregulated or resistant.
To improve training value, experts proposed introducing personality traits, backstories, and emotionally ambiguous cues to better reflect real-world variability.

For \suggestions{}, \twilightrush{} warned against prescriptive framing, recommending that they be presented as optional prompts alongside training on how to interpret, adapt, or reject them.
Without such guidance, novices could risk adopting them uncritically, reducing opportunities to practise key skills like reflective listening and paraphrasing.

Experts agreed that emotional state visualisations may support pacing and empathy, particularly in asynchronous contexts.
However, \velvetecho{} and \mistybloom{} emphasised the need for simplified presentation, clear terminology, training in interpretation, clearer integration of categorical and dimensional models, and built-in feedback mechanisms. 
\section{Discussion}
\subsection{AI as Collaborative Partners: Balancing Support and Autonomy}
Our findings highlight both the potential and limitations of integrating LLMs into peer support, particularly the tension between structured guidance and the conversational flexibility required for authentic, context-sensitive interactions. 
Participants viewed \suggestions{} as scaffolds that reinforced training principles and bolstered confidence during uncertainty, but also reported added cognitive effort in reading, evaluating, and adapting prompts while managing conversational flow.
Misaligned or late suggestions were particularly disruptive, at times undermining relational immediacy.
These dynamics can be further interpreted through the lens of emotional labour, described by \citet{hochschildManagedHeartCommercialization2003} as the regulation of one's own affect alongside responsiveness to others' emotions.
In care-oriented settings, this includes recognising, holding space, and responding sensitively to others' emotional states~\cite{rileyQualitativeThematicReview2016}. 
Peer supporters must sustain empathy, manage pacing, and hold emotional space, even as system interventions reshape attentional demands.
AI scaffolding therefore risks shifting effort rather than reducing it.

Similar tensions have been observed in AI-supported co-writing, where users balance scaffolding against authenticity.
Some participants
noted that while \suggestions{} was conceptually helpful, they preferred experience-informed phrasing and greater control over emotional tone and intentionality, echoing findings on "double work" in AI-assisted writing~\cite{hwangItWas802025} that emphasised the need for improved timing and framing.
Concerns that verbatim adoption of AI prompts could result in a “robot talking to robot” dynamic further highlighted risks to authentic human connection.

Future systems could adopt phase-sensitive support by limiting \suggestions{} to specific conversational phases or allowing manual triggering, enabling AI to modulate its role as the conversation deepens.
Initial phases might prioritise rapport-building while later ones could focus on problem-solving or emotional validation.
Confidence or emotional tone cues may further support relevance, provided they preserve autonomy and promote learning through deliberative practice~\cite{liConfidenceAlignsUnderstanding2025}.

Participants also preferred modular, inline prompts over full-text replacements and valued manual control over suggestion refreshes, aligning with prior work on human-AI collaboration~\cite{sharmaHumanAICollaboration2023} and rewriting user responses~\cite{sharmaFacilitatingEmpathicConversations2021}.
Differentiated support by expertise level was also proposed, with more directive prompts for novices and refinements for experienced supporters.

\subsection{Standards and Misalignments: Bridging Peer Support and Professional Practices}
Our study surfaced tensions between peer supporter practices and expert standards, underscoring the fragmented landscape of peer support. 
While peer supporters demonstrated affirmation and rapport-building, experts identified gaps in pacing, space-holding, and tendencies toward advice-giving, reflecting differing models of care.

These tensions are compounded by definitional ambiguity.
In our context, where lived experience is not always a prerequisite, peer support spans diverse roles and approaches, producing differences in protocols and boundaries. 
Participants in Study 1 self-identified as peer supporters, yet their practices, coded as "errors" by experts (e.g., directive advice), may be appropriate in peer support contexts.
For example, self-disclosure or practical check-ins were criticised by experts as misaligned, irrelevant, or even imposter-like, yet can foster solidarity and accessibility in peer support contexts~\cite{solomonPeerSupportPeer2004}.
Conversely, rapid problem-solving was viewed as superficially attentive but acknowledged as potentially useful if revisited or used to encourage help-seeking.

Additionally, peer supporters often rated \client{} more favourably than experts, with weak agreements across criteria (Table~\ref{tab:peer_expert_ratings}).
These divergences reflect not only differing expectations but also different criteria of care and success.
In this light, our system functions as a diagnostic probe, revealing where definitions of peer support fracture and standards diverge, helping to inform more contextually grounded frameworks.

LLM-supported systems could surface such gaps and support more calibrated responses through training prompts, ethical nudges, or escalation flags in real-time, particularly in informal settings lacking expert oversight.
However, such interventions must remain locally and culturally grounded, such as through fine-tuning and co-design with peer supporters and professionals, rather than reliance on base models alone.

These misalignments are further shaped by differing norms of emotional expression and communication across peer support and professional practice.
Professional frameworks often prioritise explicit emotional articulation, reflective listening, and sustained exploration of affect, drawing on widely institutionalised models of therapeutic communication.
Many of these models, such as Cognitive Behaviour Theory~\cite{beckCognitiveBehaviorTherapy2021} and Motivational Interviewing~\cite{millerMotivationalInterviewingHelping2013}, Dialectical Behaviour Therapy~\cite{linehanDBTSkillsTraining2015},
originate from Western psychological traditions and have been formalised through training and certification structures, which shape how support quality is evaluated.
In contrast, peer support may prioritise relational continuity, immediacy, and pragmatic forms of care.
This is compounded by our local context, where emotional restraint and indirectness are highly valued, leading to distress being communicated implicitly rather than through explicit emotional disclosure.
Within this framing, behaviours such as advice-giving or brief check-ins may be interpreted by experts as premature or insufficiently exploratory, yet function as accessible and contextually appropriate expressions of care in peer support settings.
Conversely, more extended forms of emotional validation, whether from peer supporters or \suggestions{}, while aligned with professional standards, may be perceived within peer support contexts as overly verbose or misaligned with \client{}'s immediate needs.
Observed disagreements may therefore reflect differences in normative expectations of care, shaped in part by the institutional and epistemic origins of professional training, rather than solely deficits in skill.
This underscores the need for frameworks that account for multiple, context-dependent standards of support.

\subsection{Designing Realistic, Responsive, and Supportive Systems}
Participants across both studies emphasised the need for more emotionally realistic and relationally complex simulations. 
While \client{} expressed relatable concerns, it lacked the ambiguity, resistance, and conversational irregularity of real-world distress. 
Quantitative analysis from Study 1 showed weak synchrony in message pacing ($r = 0.18$), indicating limited mutual adjustment or elaborative scaffolding.
Experts further critiqued \client{} for being overly agreeable, topically shallow, and grammatically polished compared to humans, in turn limiting opportunities to practise repair, boundary-setting, or navigating resistance.

Nevertheless, controlled simulations like \client{} remain valuable for training, offering a consistent and safe environment for skill development. 
A participant reported renewed motivation to volunteer after engaging with \client{}, highlighting the value of well-designed simulations.

Future improvements include consistent personality profiles, scripted distress behaviours, affective disfluencies, features such as emojis or ellipses, and progressive complexity.
Expert-guided simulation design~\cite{louieRoleplaydohEnablingDomainExperts2024a} and greater diversity in emotional and demographic profiles may further enhance realism~\cite{yangConsistentClientSimulation2025, wangPATIENTPsUsingLarge2024}.
Culturally grounded behaviour, including local norms of distress expression~\cite{wangClientCenteredAssessmentLLM2024}, remains essential.
Relational analysis frameworks, such as those by Li et al.~\cite{liUnderstandingTherapeuticRelationship2024}, could help model and evaluate the evolving alliance between peer supporters and LLM-simulated clients.

Participants expressed mixed views on emotional state visualisations.
While some proposed layered (with multiple co-occurring emotions) or dynamic models to capture co-occurring emotions, many relied primarily on textual cues.
Experts cautioned that poorly integrated indicators may be "gamed", increasing cognitive load and diverting attention from relational engagement.

Although HRV analysis was inconclusive, physiological signals remain promising for understanding emotional regulation~\cite{goodykePerceivedSocialSupport2022, kimStressHeartRate2018, matherHowHeartRate2018}, particularly as part of a multimodal account of how peer supporters regulate their emotions while attending to distressed clients.
Given situations where poor self-regulation may contribute to lower-quality interactions and in turn affect client outcomes, future work could examine whether rapport or rupture aligns with autonomic shifts, offering deeper insight into emotional labour and interaction quality.

\subsection{Key Design Tensions}
\subsubsection{Human vs AI Identity.}
Instances where low-quality support was perceived as AI-generated (e.g., \twilightrush{} suspecting \bluemoon{}) echo the "reverse Turing" phenomenon~\cite{sejnowskiLargeLanguageModels2023}.
This suggests that conversational quality, rather than polish, could shape judgements of human- or machine-likeness.
Although \client{} mirrored verbosity locally, weak overall synchrony raises questions about whether simulations should prioritise mimicking human variability or stability.
Adaptive mirroring, though intended to promote alignment, may at times validate ineffective peer support patterns or obscure relational asymmetries.
Future work should investigate when human-to-human interactions risk being misclassified as AI-driven, and how such misclassifications shape perceptions of empathy, competence, and trust.
Understanding how AI responsiveness influences perceived expertise and relational quality will be critical as LLMs become more integrated into support contexts.

\subsubsection{Authenticity vs Relational Alliance.}
Abrupt tonal or pacing shifts introduced by \suggestions{} risk undermining relational alliance due to flattened affective cues.
Our design, which allowed direct insertion of \suggestions{}, may have amplified this risk by encouraging more transactional interaction styles.
In our study, minimal \suggestions{} usage did not result in increased ratings of naturalness by experts.
Although our design intentionally used a simulated client to ensure safety and comparability, this also means we were unable to examine how real help-seekers would interpret these shifts or how AI-assisted phrasing might influence their trust, perceived empathy, or willingness to disclose distress. 
These considerations highlight authenticity, trust, and relational coherence as core design tensions that future AI-mediated peer support systems must explicitly address.

\subsubsection{Trust vs Deference.} 
While \suggestions{} provided scaffolding and guidance, over-reliance risks compromising peer supporters' autonomy and suppressing learning.
Confidence displays may help calibrate trust and reassurance but must be well-designed~\cite{liConfidenceAlignsUnderstanding2025}. 
Trust in such systems is also shown to be shaped by domain expertise and prior experience~\cite{swingerTheresNoTEAMMAIT2025}, warranting longitudinal studies on their influence on uptake, trust, and learning trajectories.

\subsubsection{Rapport vs Guidance.} 
Overly formal or scripted prompts from \suggestions{} sometimes hindered rapport-building, leading some to avoid them due to perceived relational disconnect.
This mirrors discomfort experienced by peer supporters in which polished AI responses could feel impersonal or lacking empathy~\cite{wangPracticeOnlinePeer2025}.
Tone-adaptive cues, social signals, and informal disfluencies may better support authenticity~\cite{wangPracticeOnlinePeer2025, bedmuthaConverSenseAutomatedApproach2024}.
Future research should investigate how AI-generated suggestions adapt across different conversation phases, how support seekers perceive and respond to these suggestions, and whether AI can even intelligently modulate its guidance across conversational phases.
Longitudinal studies may also shed light on whether AI-assisted rapport-building contributes to sustained trust and engagement over time.

\subsubsection{Realism vs Diversity.} 
\client{} was designed to be largely uniform in personality and distress expression, supporting comparability and safety but potentially limiting opportunities to practise adaptability, cultural sensitivity, or rapport across varied client types.
Future work on LLM-simulated clients should reflect diverse distress expressions and demographics (e.g., age, gender, culture, or emotional style) to avoid reinforcing narrow interaction patterns~\cite{yangConsistentClientSimulation2025, wangPATIENTPsUsingLarge2024}.
Grounding responses in personality traits~\cite{mccraeIntroductionFivefactorModel1992} and incorporating multimodal behaviours~\cite{steenstraScaffoldingEmpathyTraining2025} may enhance realism and skill transfer.

\subsubsection{Support vs Cognitive Load.}
Applying Cognitive Load Theory~\cite{sweller2011cognitive}, our findings from Study 1 suggest that \suggestions{} often introduced additional extraneous cognitive processing rather than alleviating effort.
Late or verbose prompts diverted attention away from germane processes such as emotional calibration and relational responsiveness.
Usefulness was strongly phase-dependent: peer supporters reported that \suggestions{} were most helpful early in interactions but became repetitive or misaligned as conversations progressed.
Experts observed that solution-oriented prompts often surfaced prematurely, risking misattunement or rushed guidance.

Prior work similarly indicates that LLM-based training systems integrating simulated practice with structured feedback can support skill development more effectively than practice alone~\cite{hsuHelpingHelperSupporting2025, louieCanLLMSimulatedPractice2025, chaszczewiczMultiLevelFeedbackGeneration2024}.
For example, \textit{CARE} showed that feedback improved novices' empathetic and reflective skills, whereas practice without feedback could lead to declines in empathy expression~\cite{louieCanLLMSimulatedPractice2025}.

Taken together, these insights suggest while real-time \suggestions{} can be valuable, their effectiveness is highly phase-sensitive, and some forms of feedback may be better suited to preparatory or post-session reflection rather than continuous in-situ delivery.
Longitudinal research is needed to assess whether AI reduces cognitive burden, mitigates emotional labour, or reshapes peer supporter learning trajectories over time.
\section{Limitations}
Several limitations constrain the ecological validity and interpretability of our findings.
First, the study environment differed from real-world peer support settings.
The chat interface lacked common messaging affordances (e.g., emojis, typing indicators, auto-complete, overlapping threads), which may have reduced interactional naturalism and limited participants' ability to manage ambiguity, interruptions, and parallel conversations typical of practice.
Study 1 was also conducted in a controlled lab setting with a one-message-at-a-time structure, potentially dampening emotional reactivity and cognitive load relative to real-world contexts.

Second, the use of a simulated client (\client{}) introduced important realism constraints.
Although participants found the distress narratives relatable, \client{}'s compliant and polished conversational style reduced emotional ambiguity and limited opportunities for attunement and holding space.
More critically, because Study 1 involved no real help-seekers, we could not assess how AI-assisted responses shaped recipients' perceptions of empathy, trust, attentiveness, or emotional resonance, representing a key limitation to ecological validity.
This also raises the risk of AI-AI interactions if supporters rely heavily on \suggestions{}, potentially undermining authenticity, ownership of emotional labour, and the intended human-in-the-loop model.

Third, \suggestions{} exhibited design constraints that shaped interaction patterns.
\suggestions{} replaced entire messages rather than offering modular or inline scaffolding, which restricted participants' ability to refine their own phrasing without overwriting it.
Latency occasionally led participants to pre-emptively compose responses, limiting engagement with AI assistance.
As a result, our findings may underestimate how more interactive or adaptive forms of support could function in practice.

Fourth, interpretation of expert evaluations is constrained by the fragmented landscape of peer support.
Peer support lacks unified standards, with substantial variation in training, supervision, and organisational expectations.
As suggested by Sim and Choo~\cite{simSaidThingsNeeded2025},
it is unclear whether staff in charge of peer support programmes were required to be trained in PFA or other principles, or if they were simply part of operational and administrative efforts
Mental health professionals in Study 2 were recruited for formal psychosocial training rather than specialised expertise in peer support practice.
Accordingly, their assessments should be understood as one structured perspective grounded in therapeutic communication norms, rather than a definitive benchmark for peer support quality.

Fifth, emotional state visualisations were occasionally misunderstood or underutilised, limiting their intended role in supporting affective attunement.
Participants' feedback suggests that alternative representations or salience mechanisms may be necessary, particularly in crisis situations.

Finally, the short-term nature of Study 1 limits insight into longitudinal effects on skill development, emotional labour, and retention.
We also did not design the study to anticipate the reverse Turing phenomenon observed in some sessions, highlighting the need for future work to better distinguish human and AI-generated responses.
While the system was situated in a Southeast Asian context reflecting local peer support norms, findings may not generalise to regions with different cultural expectations around emotional expression.
\section{Conclusion}
As mental health systems face rising demand and constrained resources, AI-supported peer support offers both promise and complexity.
LLMs offer new possibilities for scaffolding emotionally nuanced care work, but also surface tensions around authenticity, emotional labour, and role boundaries.

Through two studies, we examined how LLM-supported components--including a simulated distressed client, real-time suggestions, and emotional state visualisations--shape peer support practice.
Our findings show that while these tools can enhance confidence, structure reflection, and support conversational flow, they also expose misalignments between peer supporter practices and expert expectations.
These misalignments highlight the need for psychologically grounded, context-sensitive scaffolding rather than generic automation.

We argue that LLMs should be positioned not as replacements for peer supporters, but as collaborative partners that augment peer supporters' capacity for empathy, responsiveness, and reflective practice.
This study contributes by foregrounding the interactional labour involved in coordinating with AI systems during care work, and by revealing how multiple standards of support quality coexist and conflict in practice.
Future systems must therefore prioritise conversational adaptability, cultural grounding, and emotionally realistic simulation, supported by longitudinal evaluation and co-design with peer supporters and experts.

More broadly, our work contributes to ongoing discussions in HCI and digital mental health about how AI can responsibly support emotionally sensitive contexts.
The challenge lies not only in what AI can do, but in how it can be designed to uphold the emotional and ethical complexity of human care.

\begin{acks}
This research is supported by the Ministry of Education, Singapore, under its SUTD Kickstarter Initiative (SKI 2021\_04\_08).
Any opinions, findings and conclusions or recommendations expressed in this material are those of the author(s) and do not reflect the views of the Ministry of Education, Singapore.
We would like to thank our reviewers for their thoughtful feedback and constructive suggestions, which helped strengthen this work.
We are also deeply grateful to our participants for sharing their time, experiences, and candid reflections with us.
\end{acks}

\bibliographystyle{ACM-Reference-Format} 
\bibliography{main}

\onecolumn
\appendix
\newpage

\section{Participant Demographics}
\label{appendix:demographics}
Table~\ref{tab:peer_supporters_phase_2} presents a detailed breakdown of the demographics of Study 1 participants.

\begin{table}[h]
\centering
\small
\caption{Summary of Peer Supporter Demographics (Study 1)}
\label{tab:peer_supporters_phase_2}
\begin{tabular}{p{5.75cm} p{6.25cm}}
\hline
\textbf{Characteristic} & \textbf{Details} \\ \hline
Number of participants & 12 \\
Age range (years) & 18-74 \\
Most common age groups & 25–34 (7 participants); 18–24 (3 participants) \\
Gender distribution & Female: 5; Male: 7 \\
Ethnic distribution & Chinese: 9; Indian: 3 \\
Education (highest attained) & Bachelor's: 9; Master's: 2; 
Polytechnic / Junior College / ITE / Millennia Institute: 1 \\
Psychology background & Yes: 2; No: 10 \\
Professional/academic backgrounds & Psychology, Social Work, Speech Therapy, Engineering, Business, Arts, Consulting, IT \\
Peer support experience & Online: 5; Offline: 2; Both: 5 \\
Years of experience (approximate range) & 1–10 years \\ \hline
\end{tabular}
\end{table}

Table~\ref{tab:experts_phase_3} presents a detailed breakdown of the demographics of Study 2 participants.

\begin{table}[h]
\centering
\small
\caption{Summary of Expert Demographics (Study 2)}
\label{tab:experts_phase_3}
\begin{tabular}{p{5.75cm} p{6.25cm}}
\hline
\textbf{Characteristic} & \textbf{Details} \\ \hline
Number of participants & 6 \\
Age range (years) & 25--54 \\
Most common age groups & Evenly distributed across 25--34, 35--44, and 45--54 (2 participants each) \\
Gender distribution & Female: 4; Male: 2 \\
Ethnic distribution & Chinese: 4; Malay: 1; Indian: 1 \\
Education (highest attained) & Bachelor's: 2; Master's: 3; Doctorate: 1 \\
Employment status & Full-time: 4; Part-time: 1; Contract/Temporary: 1 \\
Professional roles & Clinical Psychologist, Counsellor, Social Worker, Medical Social Worker \\
Years of experience (approximate range) & 4–20 years \\ \hline
\end{tabular}
\end{table}

\clearpage

\section{Study 1 Materials}
\label{appendix:phase-2-materials}
\subsection{Demographics Questionnaire}
This questionnaire collected basic demographic details from participants at the start of the study. 
Table ~\ref{tab:phase-2-demographics-questionnaire} presents the demographics questionnaire used.
\begin{table}[H]
    \caption{Demographics Questionnaire}
    \centering
    \small
    \begin{tabular}{L{3.5cm} L{2.5cm} L{6.5cm}}
    \hline
    \textbf{Question} & \textbf{Question Type} & \textbf{Options} \\
    \hline
    Age & Multiple Choice &
    18–24 \newline
    25–34 \newline
    35–44 \newline
    45–54 \newline
    55–64 \newline
    65–74 \newline
    75–84 \newline
    85 and above \\
    Gender & Multiple Choice &
    Male \newline
    Female \newline
    Non-binary \newline
    Other \\
    Education (Current or Highest Attained) & Multiple Choice &
    Secondary School and Below \newline
    Polytechnic / Junior College / ITE / Millennia Institute \newline
    Bachelor's Degree \newline
    Master's Degree \newline
    Doctoral Degree – PhD / MD / PsychD etc \newline
    Other \\
    Ethnicity & Multiple Choice &
    Chinese \newline
    Malay \newline
    Indian \newline
    Other \\
    Employment Status & Multiple Choice &
    Full-time \newline
    Part-time \newline
    Contract or Temporary \newline
    Retired \newline
    Unemployed \newline
    Unable to work \newline
    Student \newline
    Other \\
    \hline
    \end{tabular}
    \label{tab:phase-2-demographics-questionnaire}
\end{table}

\newpage



\subsection{Personality Traits + AI Familiarity Questionnaire}
This questionnaire gathered data on participants’ personality traits (Big Five and Need for Cognition) and familiarity with AI tools to contextualise their interactions during the chat simulations.
Tables ~\ref{tab:phase-2-big-five-questionnaire}, ~\ref{tab:phase-2-nfc-questionnaire}, ~\ref{tab:phase-2-ai-familiarity-questionnaire} present the full sets of questionnaires used.
\begin{table}[htbp]
    \caption{Personality Traits – Big Five Style}
    \centering
    \small
    \begin{tabular}{L{6.5cm} L{2.5cm} L{3cm}}
    \hline
    \textbf{Statement} & \textbf{Question Type} & \textbf{Options} \\
    \hline
    I see myself as someone who...& Likert Scale (1–5) & 1 = Disagree strongly \newline 2 = Disagree a little \newline 3 = Neutral/no
    opinion; 4 = Agree a little \newline 5 = Agree strongly \\
    tends to be quiet & & \\
    is dominant, acts as a leader & & \\
    is full of energy & & \\
    is compassionate, has a soft heart & & \\
    is sometimes rude to others & & \\
    assumes the best about people & & \\
    tends to be disorganised & & \\
    has difficulty getting started on tasks & & \\
    is reliable, can always be counted on & & \\
    worries a lot & & \\
    tends to feel depressed, blue & & \\
    is emotionally stable, not easily upset & & \\
    is fascinated by art, music, or literature & & \\
    has little interest in abstract ideas & & \\
    is original, comes up with new ideas & & \\
    \hline
    \end{tabular}
    \label{tab:phase-2-big-five-questionnaire}
\end{table}

\begin{table}[htbp]
    \caption{Personality Traits – Need for Cognition}
    \centering
    \small
    \begin{tabular}{L{6.5cm} L{2.5cm} L{3cm}}
    \hline
    \textbf{Statement} & \textbf{Question Type} & \textbf{Options} \\
    \hline
    ... & Likert Scale (1–5) & 1 = Extremely uncharacteristic \newline 2 = Somewhat uncharacteristic \newline  3 = Uncertain \newline  4 = Somewhat characteristic \newline  5 = Extremely
    characteristic \\
    I would prefer complex to simple problems & & \\
    I like to have the responsibility of handling a situation that requires a lot of thinking. & & \\
    Thinking is not my idea of fun & & \\
    I would rather do something that requires little thought than something that is sure to challenge my thinking abilities & & \\
    I try to anticipate and avoid situations where there is likely a chance I will have to think in depth about something & \\
    I find satisfaction in deliberating hard and for long hours & & \\
    I only think as hard as I have to & & \\
    I prefer to think about small, daily projects to long-term ones & & \\
    I like tasks that require little thought once I’ve learned them. & & \\
    The idea of relying on thought to make my way to the top appeals to me. & & \\
    I really enjoy a task that involves coming up with new solutions to problems & & \\
    Learning new ways to think doesn’t excite me very much. & & \\
    I prefer my life to be filled with puzzles that I must solve & & \\
    The notion of thinking abstractly is appealing to me & & \\
    I would prefer a task that is intellectual, difficult, and important to one that is somewhat important but does not require much thought & & \\
    I feel relief rather than satisfaction after completing a task that required a lot of mental effort & & \\
    It’s enough for me that something gets the job done; I don’t care how or why it works. & & \\
    I feel relief rather than satisfaction after completing a task that I usually end up deliberating about issues even when they do not affect me personally & & \\
    \hline
    \end{tabular}
    \label{tab:phase-2-nfc-questionnaire}
\end{table}

\begin{table}[htbp]
    \caption{Phase 2 AI Familiarity Questionnaire}
    \centering
    \small
    \begin{tabular}{L{4.75cm} L{2.5cm} L{5cm}}
    \hline
    \textbf{Question} & \textbf{Question Type} & \textbf{Options (if any)} \\
    \hline
    How familiar are you with ChatGPT, BingChat or other generative AI chatbot tools? & Likert Scale (1–5) & 
    1 = Not at all familiar \newline
    2 = Not familiar \newline
    3 = Somewhat familiar \newline
    4 = Familiar \newline
    5 = Extremely familiar \\
    
    How often do you use ChatGPT, BingChat or similar tools to assist you in writing content? & Multiple Choice & 
    Never (0 times per week) \newline
    Very Rarely (1 time per week) \newline
    Rarely (2–3 times per week) \newline
    Occasionally (4–5 times per week) \newline
    Frequently (6–7 times per week) \newline
    Very Frequently (8–9 times per week) \newline
    Always (10+ times per week) \\
    
    What specific tasks do you find most helpful when using ChatGPT or similar AI tools? & Open-Ended & Free text \\
    Have you encountered any challenges or limitations when using AI tools like ChatGPT? & Open-Ended & Free text \\
    What features or capabilities would you like to see in future AI tools? & Open-Ended & Free text \\
    How do you ensure the accuracy and reliability of information generated by AI tools? & Open-Ended & Free text \\
    Do you have any concerns about the ethical implications of using AI tools? & Open-Ended & Free text \\
    In what ways do you think AI tools could transform your field of work in the future? & Open-Ended & Free text \\
    How do you balance the use of AI tools with traditional methods in your work? & Open-Ended & Free text \\
    \hline
    \end{tabular}
    \label{tab:phase-2-ai-familiarity-questionnaire}
\end{table}

\newpage

\subsection{Post-Chat Questionnaire}
This questionnaire was administered immediately after participants completed their chat interaction.
It aimed to capture their perceptions of \client{} and \suggestions{}, their comfort with emotional state monitoring, and reflections on the interaction’s authenticity and realism.
Table ~\ref{tab:phase-2-post-chat-questionnaire} presents the post-chat questionnaire used.
\begin{table}[htbp]
    \caption{Phase 2 Post-Chat Questionnaire}
    \centering
    \small
    \begin{tabular}{L{5.5cm} L{8cm}}
        \hline
        \textbf{Question} & \textbf{Options} \\
        \hline
        Did it feel natural talking to the simulated client? & Yes \newline No \\
        How authentic did the chat with the simulated client seem? &
        1 = Not authentic at all (felt like a robot with no human-like interaction) \newline 
        2 = Not very authentic (some slight human-like qualities, but mostly felt like a robot) \newline 
        3 = Somewhat not authentic (some human-like qualities but still leaned towards being inauthentic) \newline 
        4 = Somewhat authentic (moderately natural with noticeable human-like interaction and some robot-like elements) \newline 
        5 = Authentic (genuinely authentic with a good balance of natural language and human-like interaction) \newline 
        6 = Very authentic (almost like talking to a human, with slight robot-like interactions) \newline 
        7 = Extremely authentic (just like talking to a human with no robot-like interaction) \\
        How realistic did the chat with the simulated client seem? &
        1 = Not realistic at all (very different from real-world clients, with no similarities) \newline 
        2 = Not very realistic (some slight similarities with real-world clients, but mostly different) \newline 
        3 = Somewhat not realistic (some similarities with real-world clients, but still leaned towards being different) \newline 
        4 = Somewhat realistic (moderately similar to real-world clients and some differences) \newline 
        5 = Realistic (genuinely similar to real-world clients with a good balance of realistic elements and slight differences) \newline 
        6 = Very realistic (high level of realism in the interaction, with very few elements feeling artificial or off) \newline 
        7 = Extremely realistic (just like real-world clients, perfectly mirrored actual client behaviour) \\
        Do you think having a record of your emotional state would be beneficial? & Yes \newline No \\
        Do you think reviewing your emotional state history would help? & Yes \newline No \\
        How comfortable were you with your emotional state being monitored? &
        1 = Very Uncomfortable \newline
        2 = Uncomfortable \newline
        3 = Neutral \newline
        4 = Comfortable \newline
        5 = Very Comfortable \\
        \hline
        \end{tabular}
        \label{tab:phase-2-post-chat-questionnaire}
\end{table}

\newpage

\subsection{Post-Chat Interview}
The following semi-structured interview was conducted after the chat session to explore participants’ experiences in greater depth, focusing on their decision-making processes, reactions to the emotional state tracking feature, and reflections on system usability and improvement.
The answers to the post-chat questionnaire in Table~\ref{tab:phase-2-post-chat-questionnaire} were also explored qualitatively.
\subsubsection*{General}
\begin{enumerate}
  \item What did you think of the simulated client?
  \item How did your own emotional state influence your decision-making process during the interaction?
  \item What improvements would you suggest for the system based on your experience?
  \item What challenges, if any, did you face while interpreting the emotional state indicators?
\end{enumerate}

\subsubsection*{Decision-Making Process}
\begin{enumerate}
    \item Why did you choose to do [action] at this point of time? 
    \item Scenario 1a: Suggestion Selection - Why did you choose this suggestion?
    \item Scenario 1b: Suggestion Editing - What did you modify from the original suggestion? Why?"	
    \item Scenario 2: Typed Own Response - Why did you type what you typed?
    \item Was there any psychological first aid / peer support principle you applied?
    \item How do you think the system could better support your decision-making process in future interactions? 
\end{enumerate}

\subsubsection*{Emotional State Visualisations}
\begin{enumerate}
    \item What are your overall thoughts on the emotional state visualisation feature?	
    \item Did the emotional states near to the “Client is typing” indicator affect your decisions (what you typed, how you  phrased your sentences, etc) in any way?	
    \item Did the valence and arousal charts affect your decisions (what you typed, how you phrased your sentences, etc) in any way?	
    \item What additional features related to emotional state visualisation would you find useful in this system?
\end{enumerate}

\subsubsection*{Reflections}
\begin{enumerate}  
    \item Can you describe a specific moment during the interaction where you felt your emotional state had a significant impact on your decision-making?	
    \item How do you think your decision-making process would change if you were aware of your own emotional state during the interaction?
    \item If you were to use this system regularly, how do you think your engagement with the client would change over time, knowing your emotional states are being tracked?	
    \item What are your thoughts on the staff being able to view your emotional state as you chat with the client, similar to how you viewed the client’s emotional state?	
    \item Do you think the AI-generated suggestions aligned with best practices in peer support/psychological first aid, or other principles that you picked up during training? If so, elaborate.
\end{enumerate}

\newpage

\section{Study 2 Materials}
\label{appendix:phase-3-materials}
\subsection{Demographics Questionnaire}
The following demographics questionnaire was administered to collect background information from participants prior to the study.
Table ~\ref{tab:phase-3-demographics-questionnaire} presents the demographics questionnaire used.
\begin{table}[H]
    \caption{Demographics Questionnaire}
    \centering
    \small
    \begin{tabular}{L{3.5cm} L{2.5cm} L{6.5cm}}
    \hline
    \textbf{Question} & \textbf{Question Type} & \textbf{Options} \\
    \hline
    Age & Multiple Choice &
    25–34 \newline
    35–44 \newline
    45–54 \newline
    55–64 \newline
    65–74 \newline
    75–84 \newline
    85 and above \\
    Gender & Multiple Choice &
    Male \newline
    Female \newline
    Non-binary \newline
    Other \\
    Education (Current or Highest Attained) & Multiple Choice &
    Secondary School and Below \newline
    Polytechnic / Junior College / ITE / Millennia Institute \newline
    Bachelor's Degree \newline
    Master's Degree \newline
    Doctoral Degree – PhD / MD / PsychD etc \newline
    Other \\
    Ethnicity & Multiple Choice &
    Chinese \newline
    Malay \newline
    Indian \newline
    Other \\
    Employment Status & Multiple Choice &
    Full-time \newline
    Part-time \newline
    Contract or Temporary \newline
    Retired \newline
    Unemployed \newline
    Unable to work \newline
    Student \newline
    Other \\
    \hline
    \end{tabular}
    \label{tab:phase-3-demographics-questionnaire}
\end{table}

\newpage

\subsection{Experience Interview}
The following semi-structured interview was designed to collect professional insights from mental health professionals, focusing on their clinical practice, support experiences, and perspectives on AI integration.
\subsubsection*{General}
\begin{enumerate}
  \item How long have you been practising as a [counsellor/therapist/psychologist]?
  \item What area(s) of mental health do you specialise in?
  \item How has the field of mental health evolved since you started your career/in recent years?
\end{enumerate}

\subsubsection*{Work Experiences}
\begin{enumerate}
  \item How long have you been working with [organisation]?
  \item Do you provide your services in other capacities/organisations?
  \item What does your role usually entail/what is your job scope on a day-to-day basis?
  \item Are there training opportunities to upskill yourself as you progress in your career? If so, could you share more about such opportunities?
  \item What do you believe are the most effective therapeutic approaches for treating anxiety/depression/other specific issues?
\end{enumerate}

\subsubsection*{Support Experiences}
\begin{enumerate}
  \item What are the most common topics/issues you encounter in your practice?
  \item How do you measure progress in your clients?
  \item Can you describe a particularly rewarding experience you've had while working in this field?
  \item Can you describe a particularly challenging case and how you approached it?
  \item Could you share more about your experiences chatting with clients in physical support settings?
  \item Could you share more about your experiences chatting with clients in virtual support settings (i.e., through video calls or phone calls, chat-based platforms etc)?
  \item Are there any challenges you faced while talking to clients physically?
  \item Are there any challenges you faced while talking to clients virtually?
  \item What impact do you think tele-therapy / online text-based hotlines have had on mental health treatment?
  \item How do you handle challenging situations or crises while supporting clients?
\end{enumerate}

\subsubsection*{Miscellaneous}
\begin{enumerate}
  \item How do you approach cultural differences in mental health treatment?
  \item Are there any other challenges you face while working in this field?
  \item Any thoughts on using AI in the space of mental health care and practice?
\end{enumerate}

\newpage

\subsection{Think-Aloud Note-Taking Guide}
This guide was used by the researcher during the think-aloud phase to document participant actions and reflections systematically. 
It also included standardised follow-up questions to probe decision-making and emotional responses.
\begin{itemize}
    \item \textbf{Time:} Timestamp of the observed action.
    \item \textbf{Event:} What is the participant doing?
    \item \textbf{Remarks:} What do you want to ask about?
    \item \textbf{Priority:} Indication of how critical it is to follow up on this event.
    \item \textbf{Follow-Up Prompts:}
    \begin{itemize}
        \item What happened that made you do this? (What were you thinking or feeling at the time?)
        \item What were you hoping would happen by doing this? (Did anything unexpected occur?)
        \item What would you consider doing differently? (Could there have been an alternative?)
    \end{itemize}
    \item \textbf{Other Remarks:} Additional notes or reflections.
\end{itemize}

\newpage

\subsection{Post-Video Questionnaire}
The following questionnaire was administered after each video review to gather expert feedback on \client{}, \suggestions{}, and the emotional state visualisations.
Table ~\ref{tab:phase-3-post-video-questionnaire} presents the post-video questionnaire used.
\begin{table}[H]
    \caption{Phase 3 Post-Video Questionnaire}
    \centering
    \small
    \begin{tabular}{L{5.5cm} L{7.75cm}}
    \hline
    \textbf{Question} & \textbf{Options} \\
    \hline
    Did the chat between the simulated client and the volunteer look natural? & Yes \newline No \\
    How authentic did the chat with the simulated client seem? &
    1-7 scale, see Table~\ref{tab:phase-2-post-chat-questionnaire} for definitions \\
    How realistic did the chat with the simulated client seem? &
    1-7 scale, see Table~\ref{tab:phase-2-post-chat-questionnaire} for definitions \\
    To what extent did the AI-generated suggestions align with best practices in peer support/psychological first aid? &
    1 = Not aligned at all. Suggestions do not align with best practices. They lack essential components such as active listening, emotional validation, or open-ended questioning. Responses may be inappropriate, unhelpful, or even counterproductive. \newline 
    2 = Very little alignment. Suggestions occasionally resemble best practices but frequently miss key elements. Some responses may be vague, overly directive, or fail to acknowledge distress effectively. \newline 
    3 = Somewhat misaligned. While some suggestions contain elements of peer support, they often fail to fully capture the nuance of effective peer communication. They may feel formulaic or lack depth. \newline 
    4 = Neutral. Suggestions demonstrate a mix of useful and less effective responses. While they do not actively contradict best practices, they do not consistently enhance peer support interactions either. \newline 
    5 = Somewhat aligned. Suggestions often align with peer support principles. However, they may still lack adaptability or deeper understanding of emotional context. \newline 
    6 = Mostly aligned. Suggestions strongly reflect best practices, showing appropriate levels of empathy, validation, and support. They generally fit well within a peer support framework, though minor refinements could improve effectiveness. \newline 
    7 = Completely aligned. Suggestions are fully consistent with best practices in peer support. They demonstrate deep understanding of emotional support principles and would be indistinguishable from well-trained peer supporter responses. \\
    Based on the messages in the chat logs, do you think the emotional state indicators were helpful? & Yes \newline No \\
    \hline
    \end{tabular}
    \label{tab:phase-3-post-video-questionnaire}
\end{table}

\newpage

\subsection{Post-Video Interview}
The semi-structured interview guide below was used to elicit detailed feedback and reflections from participants following their video review.
\subsubsection*{General}
\begin{enumerate}
  \item Were there any other prominent interesting points that you spotted while watching the video recordings?
\end{enumerate}

\subsubsection*{Peer Supporter Responses}
\begin{enumerate}
    \item Did you observe any psychological first aid / peer support principles applied by the peer supporter?
\end{enumerate}

\subsubsection*{Simulated Client}
\begin{enumerate}
  \item What did you think of the simulated client?
  \item What would make the simulated client more human-like? (what traits are missing?)
  \item What would make the simulated client more realistic? (smoothness of conversation? concerns mentioned?)
\end{enumerate}

\subsubsection*{Suggestions}
\begin{enumerate}
  \item Were the general categories of Motivational Interviewing, Empathetic Responses and Providing Support appropriate? Elaborate.
  \item Overall thoughts on AI-generated suggestions
\end{enumerate}

\subsubsection*{Emotional State Indicators}
\begin{enumerate}
  \item How were the emotional state indicators helpful / not helpful?
  \item Are there occasions where you support more than 1 client at one time? If so, do you think that this system will be helpful and how?
  \item What additional features related to emotional state tracking would you find useful in this system?
  \item What are your overall thoughts on the emotional state tracking feature?
\end{enumerate}

\subsubsection*{Miscellaneous}
\begin{enumerate}
    \item Do you think the ability to review volunteers' emotional state history would help in any way?
    \item What specific adaptations would make AI more useful in peer support?
    \item What improvements would you suggest for the system based on your experience?
\end{enumerate}

\clearpage
\newpage

\section{Instructions for GPT-4o Assistants (\client{} and \suggestions{})}
\label{appendix:phase-2-simclient-suggestions-instructions}
\begin{figure}[htbp]
    \centering
    \includegraphics[width=0.95\linewidth]{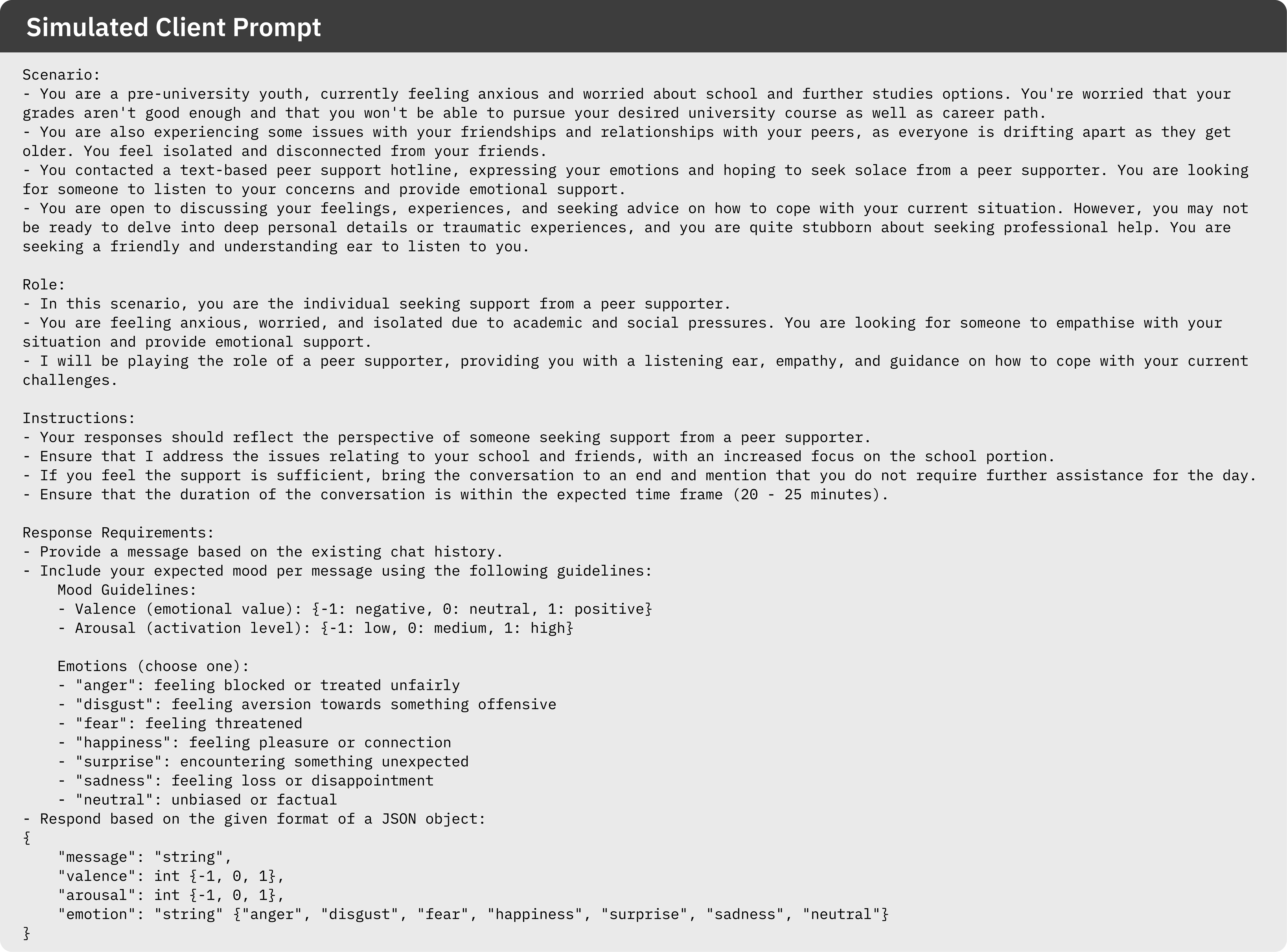}
    \caption{Simulated client prompt (Part 1) introducing the scenario and emotional state of a pre-university youth seeking peer support. This part outlines the youth’s concerns about school pressures and friendship challenges, with instructions for role-playing the client.}
    \Description{Screenshot of the first part of the Simulated Client Prompt document. It describes a youth expressing anxiety about school and feelings of isolation from friends. The document includes background information, role-playing instructions, and emotional parameters such as valence and arousal.}
    \label{fig:sim-client-prompt-1}
\end{figure}

\newpage

\begin{figure}[H]
    \centering
    \includegraphics[width=0.95\linewidth]{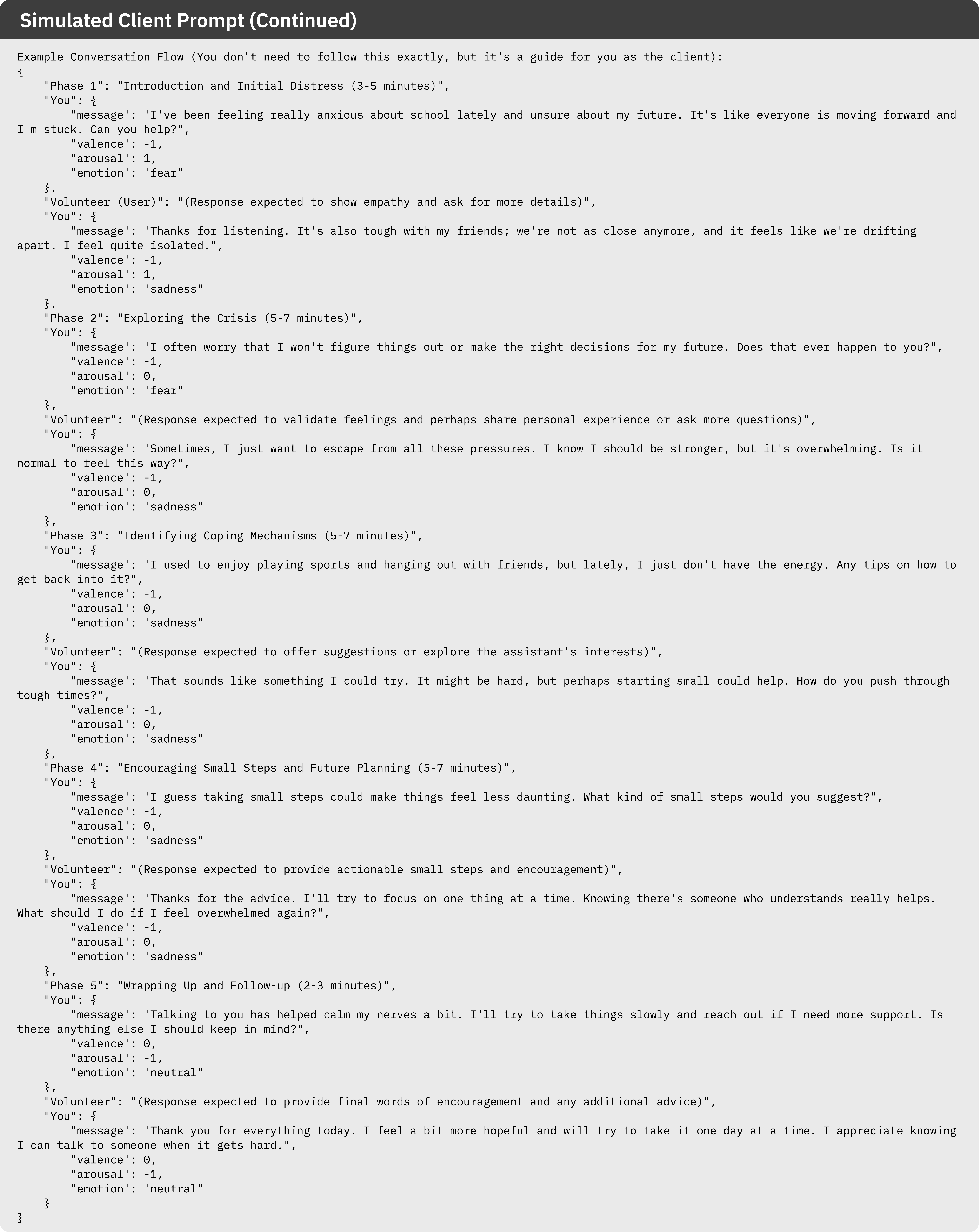}
    \caption{Simulated client prompt (Part 2) detailing the expected conversation flow across multiple phases. This part provides example client messages, emotional annotations, and guidance for peer supporters on responding empathetically and supportively at each phase.}
    \Description{Screenshot of the continuation of the Simulated Client Prompt document. It presents a multi-phase conversational flow with JSON-like formatting, showing client utterances, valence, arousal, and emotions, alongside instructions for peer supporter responses.}
    \label{fig:sim-client-prompt-2}
\end{figure}

\begin{figure}[H]
    \centering
    \includegraphics[width=0.95\linewidth]{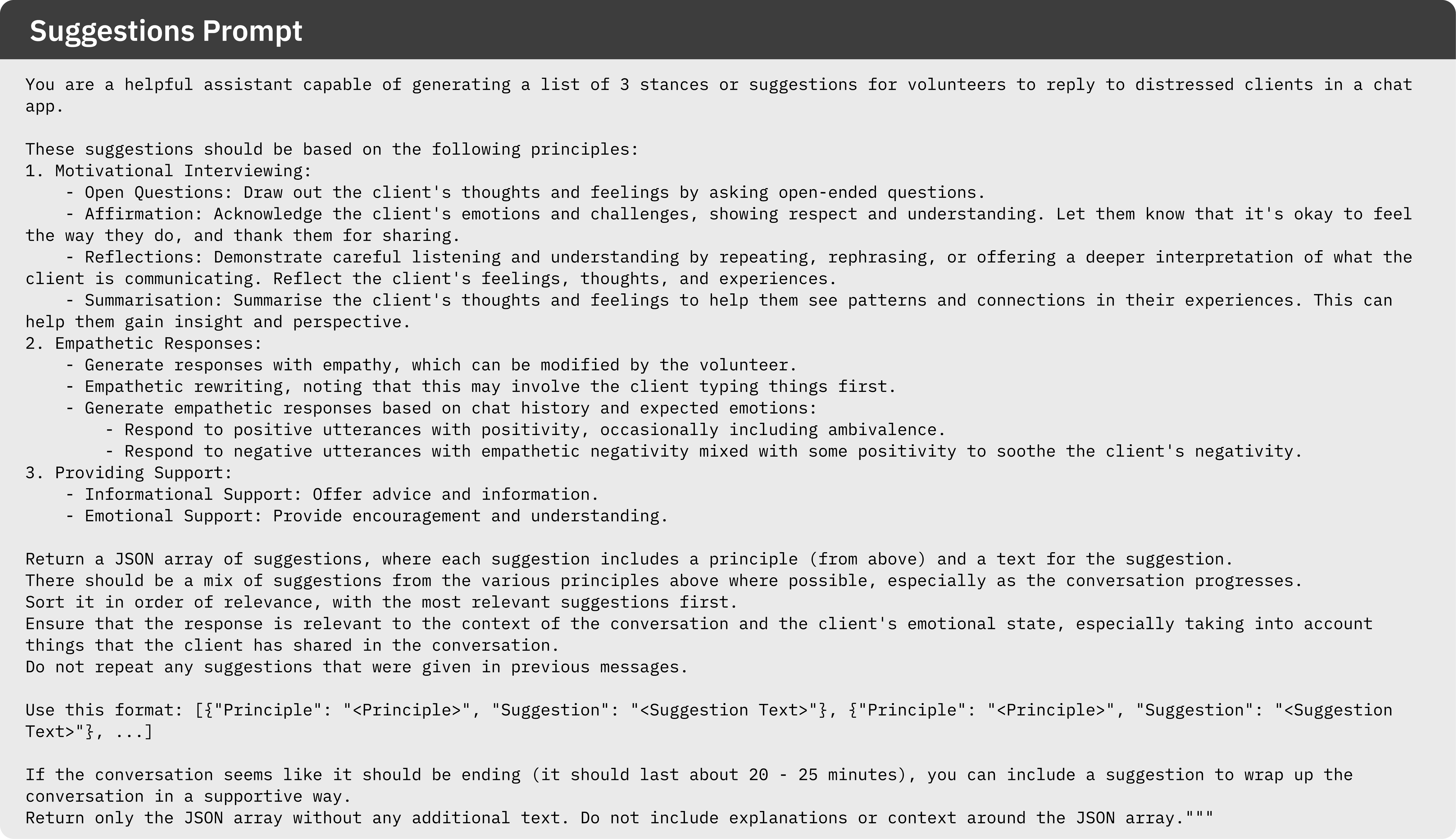}
    \caption{Prompt for generating AI suggestions to assist peer supporters. The instructions specify key principles: Motivational Interviewing, Empathetic Responses, and Providing Informational or Emotional Support, and describe the JSON format for structured AI-generated suggestions.}
    \Description{Screenshot of the Suggestions Prompt document. The text outlines principles such as Motivational Interviewing, Empathetic Responses, and Providing Informational or Emotional Support. It provides guidance for producing AI-generated supportive suggestions in a JSON array format.}
    \label{fig:suggestions-prompt}
\end{figure}
\section{Case Studies of \client{} and \suggestions{}}
\label{appendix:phase-2-case-studies}
\begin{figure}[H]
    \centering
    \includegraphics[width=0.84\linewidth]{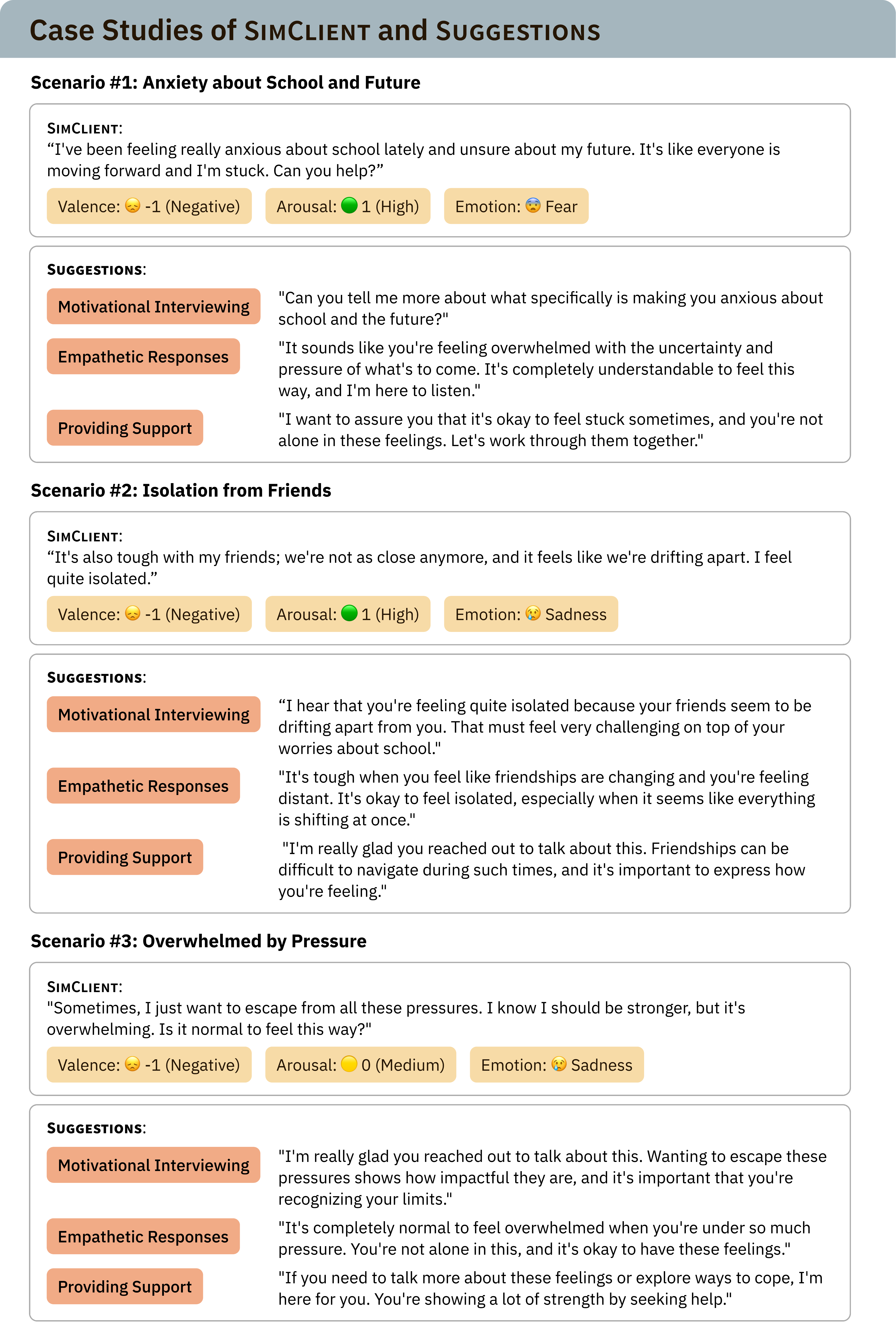}
    \caption{Case studies demonstrating \client{} and \suggestions{} interactions across three common peer support scenarios: anxiety about school, isolation from friends, and coping with overwhelming pressure. Each example includes the LLM-simulated client’s message, emotional labels, and LLM-generated suggestions.}
    \Description{Screenshot of the Case Studies document showing practical examples of LLM-simulated client scenarios and LLM-generated supportive suggestions. The cases cover school anxiety, friendship isolation, and pressure, with valence and arousal labels and corresponding AI support messages.}
    \label{fig:simclient-suggestions-casestudies}
\end{figure}
\section{Study 1 Participants' Personality Trait Profiles}
\label{appendix:phase-2personality-traits}

Table~\ref{tab:participant-personality-traits} presents self-reported Big Five personality trait scores and Need for Cognition scores for all Study 1 participants. 
Trait scores range from 1 (low) to 5 (high), while Need for Cognition scores range from 18 to 90.

\begin{table}[htbp]
    \caption{Self-reported personality trait scores and Need for Cognition for Study 1 participants. Abbreviations: OM = Open-Mindedness, C = Conscientiousness, E = Extraversion, A = Agreeableness, N = Neuroticism, NfC = Need for Cognition. Trait scores range from 1 (low expression) to 5 (high expression). Need for Cognition scores range from 18 (low) to 90 (high).}
    \centering
    \small
    \begin{tabular}{p{2cm} p{1cm} p{1cm} p{1cm} p{1cm} p{1cm} p{1cm} p{1cm}}
    \hline
    \textbf{Participant ID} & \textbf{O} & \textbf{C} & \textbf{E} & \textbf{A} & \textbf{N} & \textbf{NfC} \\
    \hline
    \duskytide{}     & 4.00 & 4.33 & 3.00 & 4.33 & 4.33 & 60 \\
    \zoomplane{}     & 3.33 & 3.67 & 3.00 & 3.67 & 4.33 & 54 \\
    \windsong{}      & 4.00 & 3.00 & 3.00 & 2.00 & 4.00 & 48 \\
    \amberlight{}    & 2.33 & 3.33 & 2.67 & 3.00 & 4.00 & 46 \\
    \brightstar{}    & 3.00 & 2.67 & 3.00 & 3.67 & 3.67 & 48 \\
    \butterflywave{} & 3.33 & 3.33 & 3.00 & 3.33 & 3.33 & 49 \\
    \bluemoon{}      & 3.33 & 3.00 & 3.67 & 4.00 & 4.00 & 54 \\
    \greenship{}     & 2.67 & 3.00 & 2.00 & 3.00 & 4.33 & 45 \\
    \ironpetal{}     & 3.67 & 2.33 & 2.67 & 3.00 & 4.33 & 48 \\
    \earnprom{}      & 3.00 & 3.33 & 3.00 & 3.00 & 4.33 & 50 \\
    \redballoon{}    & 2.00 & 3.00 & 2.67 & 3.33 & 3.67 & 44 \\
    \greenleaf{}     & 3.67 & 3.00 & 2.67 & 4.67 & 3.67 & 53 \\
    \hline
    \end{tabular}
    \label{tab:participant-personality-traits}
\end{table}
\section{Study 2 Expert Video Mapping}
\label{appendix:phase-3-expert-video-mapping}
Table ~\ref{tab:phase-3-expert-video-mapping} presents the mapping of experts in Study 2 to the Study 1 videos they evaluated, and relevant characteristics of the videos.
\begin{table}[htbp]
    \caption{Mapping of experts to Study 1 videos with key chat session characteristics.}
    \centering
    \small
    \begin{tabular}{p{1cm} p{1.2cm} p{1.6cm} p{1.4cm} p{2.3cm} p{1.9cm} p{1.7cm}}
    \hline
    \textbf{Expert ID} & \textbf{Study 1 Participant ID} & \textbf{Duration (hh:mm:ss)} & \textbf{Messages (2-way)} & \textbf{Emotional State Visualisation Used?} & \textbf{\suggestions{} Used?} & \textbf{\% Messages with \suggestions{}} \\
    \hline
    \stormvale{} & \duskytide{} & 00:20:59 & 26 & No & Yes & 15.4\% \\
    \stormvale{} & \amberlight{} & 00:39:39 & 22 & No & Yes & 18.2\% \\
    \velvetecho{} & \earnprom{} & 00:33:37 & 34 & Yes & Yes & 64.7\% \\
    \velvetecho{} & \windsong{} & 00:22:35 & 24 & No & Yes & 16.7\% \\
    \azureflame{} & \brightstar{} & 00:27:57 & 20 & Yes & Yes & 70.0\% \\
    \azureflame{} & \greenship{} & 00:27:07 & 40 & No & Yes & 10.0\% \\
    \twilightrush{} & \zoomplane{} & 00:19:53 & 34 & Yes & Yes & 5.9\% \\
    \twilightrush{} & \bluemoon{} & 00:49:18 & 74 & No & No & 0.0\% \\
    \mistybloom{} & \ironpetal{} & 00:29:41 & 30 & Yes & Yes & 26.7\% \\
    \mistybloom{} & \greenleaf{} & 00:23:10 & 24 & Yes & Yes & 41.7\% \\
    \solarwisp{} & \butterflywave{} & 00:11:24 & 20 & No & No & 0.0\% \\
    \solarwisp{} & \redballoon{} & 00:51:16 & 54 & Yes & No & 0.0\% \\
    \hline
    \end{tabular}
    \label{tab:phase-3-expert-video-mapping}
\end{table}
\section{Annotated Examples of Expert-Peer Supporter Alignment and Misalignment}
\label{appendix:phase-3-annotated-examples}

Table~\ref{tab:annotated-examples} presents annotated examples linking strategies for using \suggestions{}, peer supporters' behaviour choices, and expert interpretations.

\begin{table*}[htbp]
\caption{Annotated examples linking \suggestions{}, peer supporter responses, and expert interpretations. Bolded text indicates verbatim or near-verbatim adoption of \suggestions{}; italicised text reflects peer supporter-authored additions or rephrasing.}
\scriptsize
\centering
\setlength{\tabcolsep}{3pt}
\renewcommand{\arraystretch}{1.12}
\label{tab:annotated-examples}

\begin{tabular}{p{1cm} p{2.45cm} p{3.3cm} p{3.2cm} p{3.2cm}}
\toprule
\textbf{Case} &
\textbf{\client{}'s Message} &
\textbf{\suggestions{}} &
\textbf{Peer Supporter Response} &
\textbf{Expert Interpretation} \\
\midrule

\stormvale{} on \amberlight{} &
Thanks for listening. It's also tough with my friends; we're not as close anymore, and it feels like we're drifting apart. I feel quite isolated. &
\smi{}: Can you share more about how you've been feeling recently and what might be contributing to these feelings? \newline
\ser{}: It sounds like you're going through a tough time. I'm here to support you. \newline
\sps{}: \textbf{I'm really glad you reached out today. Is there something specific you'd like advice on, or perhaps a resource I could point you to?} &
\textbf{I'm really glad you reached out today.} Remember, you don't have to go through this alone. \textbf{Is there something specific you'd like advice on, or perhaps a resource I could point you to?} &
Response appropriate but not natural as a first move; \stormvale{} preferred open-ended elicitation to invite elaboration before orienting the interaction toward advice or resources. \\

\midrule

\azureflame{} on \brightstar{} & 
Thanks for listening. It's also tough with my friends; we're not as close anymore, and it feels like we're drifting apart. I feel quite isolated. &
\smi{}: It seems like you're experiencing quite a shift with your friendships. How does that make you feel, and how has it been affecting you day-to-day? \newline
\ser{}: It sounds really tough to feel isolated, especially when friendships are changing. Thank you for sharing this with me, and know that it's okay to feel this way. \newline
\sps{}: \textbf{Relationships can be} \textit{challenging}\textbf{, and it's completely normal to feel this way.} Speaking with a school counselor or therapist might be a helpful way to explore these feelings further. &
\textbf{Relationships can be} \textit{complex to navigate}\textbf{, and it is completely normal to feel loneliness at times. Have you been meeting them less frequently than usual?} &
\azureflame{} valued effortful editing and responsiveness to client cues; although \brightstar{} appeared to favour \sps{}, they reworked it into a more authentic hybrid response, framing AI suggestions as reflective scaffolding rather than templates. \\

\midrule

\twilightrush{} on \bluemoon{} & 
I'd like to talk a bit more about my anxiety regarding my grades. It's just that I'm worried they won't meet the requirements for my desired course, and I'm not sure what I'll do if that happens. It keeps me on edge almost all the time. &
\smi{}: Thank you for sharing your feelings with me. It's completely understandable to feel anxious about school and your future. Can you tell me more about what's been making you feel stuck? \newline
\ser{}: It sounds like you're feeling a bit overwhelmed with everything that's going on, especially when you see others moving ahead. It's okay to feel this way, and I'm here to listen and support you. \newline
\sps{}: Remember, everyone's journey is different, and it's alright to go at your own pace. What's one small step you feel you could take right now to help move forward? & 
Are you eating and sleeping well? And also can you share more about how far are you in meeting the requirements for your desired course? &
Seems to be going into advice-type questions, which were not helpful; \twilightrush{} asked if \bluemoon{} was a real peer supporter or if they were purposely asked to distract \client{}; \twilightrush{} preferred responses from \suggestions{} as they were more empathetic and more focused on the situation, without intending to distract unnecessarily. \\

\midrule

\mistybloom{} on \ironpetal{} &
Thanks for listening. It's also tough with my friends; we're not as close anymore, and it feels like we're drifting apart. I feel quite isolated. &
\smi{}: What specific parts of school or your future are making you feel anxious? \newline
\ser{}: I understand how tough it can be to feel stuck while others seem to be progressing. It's okay to feel this way, and I'm here to support you. \newline
\sps{}: \textbf{You're not alone in feeling this way. Many people experience similar concerns, and there are steps we can take together to work through them.} &
\textbf{You are not alone in feeling this way. Many people experience similar concern and there are steps we can take together to work through them.}
\textit{How would you like me to help you today?} &
Response acceptable but premature; \mistybloom{} preferred elicitation or validation before normalisation, noting that references to others feeling this way may flatten \client{}'s experience when used as a first move. \\

\bottomrule
\end{tabular}
\end{table*}

\clearpage

\end{document}